\documentclass[twocolumn,trackchanges]{aastex7}

\hypersetup{breaklinks=true, colorlinks=true, urlcolor={blue}, citecolor={blue}, linkcolor={blue}}

\usepackage{amsmath}
\usepackage{multirow}
\usepackage{cleveref}
\usepackage{soul}
\usepackage{subfig}

\begin{document}

\title{Irradiated Pulsar Planets and Companions as 511 keV Positron Annihilation Line Sources}

\author[0000-0002-1236-8510]{Zachary Metzler}
\affiliation{Department of Physics, University of Maryland, College Park}
\affiliation{NASA Goddard Space Flight Center}
\affiliation{Center for Research and Exploration in Space Science \& Technology II}
\email[show]{zmetzler@umd.edu}  

\author[0000-0002-9249-0515]{Zorawar Wadiasingh}
\affiliation{Department of Astronomy, University of Maryland, College Park}
\affiliation{NASA Goddard Space Flight Center}
\affiliation{Center for Research and Exploration in Space Science \& Technology II}
\email[show]{zorawar@umd.edu}

\begin{abstract}

Millisecond pulsars (MSPs) are prolific GeV $\gamma\text{-ray}$ emitters, and nearly 80\% of Fermi-LAT MSPs reside in compact binaries. \textcolor{black}{For the first time in the literature, we} demonstrate that the companions in these compact MSPs binaries are also 511 keV annihilation line emitters using {\tt MEGAlib} simulations (a high energy radiation transport software built with {\tt Geant4}) to compute the particle showers and resulting back splash emission from the pulsar irradiation. The 511 keV signal exhibits strong flux modulation and red/blueshifts associated with a binary orbit, enabling powerful coherent searches. Measuring the 511 keV emission would enable direct $\gamma\text{-ray}$ characterization of unusual pulsar exoplanets and companions, and allow one to identify the unambiguous presence of active pulsars whose beams do not intercept Earth. Intriguingly, the 511 keV flux is brightest for ultra-compact systems, against which pulsar surveys are systematically biased. These ultra-compact systems are also possibly prime LISA galactic sources. This necessitates future joint LISA-MeV $\gamma\text{-ray}$ techniques to characterize MSP binaries. These MSP binaries may also contribute to the puzzling source of the excess 511 keV photons near the galactic bulge and center.

\end{abstract}



\section{Introduction}\label{sec:Intro}

An excess of photons of energies $\sim0.5$ MeV from the galaxy is a mystery. It was initially discovered by a balloon-borne instrument in 1971, with its origin speculated to be a result of nuclear activation at $476 \pm 24$ keV or perhaps gravitationally redshifted 511 keV photons from the surfaces of neutrons stars \citep{1972ApJ...172L...1J,1973ApJ...184..103J}. Follow-up measurements by another balloon-borne instrument with improved spectral resolution in 1977 identified the excess' energy at $510.7 \pm 0.5$ keV. 
The current leading measurements of the 511 keV excess with INTEGRAL SPI still leave the source of the photons unresolved but have identified 4 components -- a central component, two central bulge components, and a disk component~\citep{Vedrenne2003,Skinner_20156,Siegert_2016,2017PhDT.......404S,2023Ap&SS.368...27S}. Similarly, the origin of the {\it Fermi} Large Area Telescope (LAT) \citep{Fermi-LAT:2009ihh} galactic center GeV $\gamma\text{-ray}$ excess is a mystery \citep{Ackermann_2017}. 
Many sources have been proposed to contribute to these excesses, including dark matter annihilation \citep{Ackermann_2017,Keith_2021,luque_2024,Berteaud_2022,Aghaie:2025dgl}, point-like X-ray binary jets \citep{10.1093/mnras/sty2135}, or an unresolved population of pulsars near the galactic center \citep{Bartels_2016}. COSI, an upcoming MeV $\gamma\text{-ray}$ explorer that will launch in 2027 \citep{tomsick_2023}, aims to resolve the 511~keV excess, and several MeV $\gamma\text{-ray}$ observatory concepts of varying capability could also settle the 511~keV excess question with estimated sensitivities ranging from $10^{-8}$ to $10^{-5} \,\, \mathrm{ph/cm^2/s}$ over their mission lifetime \citep{AMEGO,GRAMS,fleischhack2021amegoxmevgammarayastronomy,AMEGO-X,Orlando_2022,APT,2024NuScT..35..149P,GammaTPC,2025arXiv250220916L}.

Millisecond pulsars (MSPs) canonically form by accreting material from a companion star that increases the angular velocity of a neutron star (NS) \citep{1982Natur.300..728A,1983Natur.304..423H,BHATTACHARYA19911,1999A&A...350..928T}. \textcolor{black}{MSPs may also form through accretion-induced collapse of white dwarf stars \citep[WDs,][]{1991PhR...203....1B,2010MNRAS.402.1437H,Gautam_2022}}. Most MSPs are excellent GeV $\gamma\text{-ray}$ emitters \citep{3PC} from their outer magnetosphere and equatorial current sheets, with produced $\gamma\text{-ray}$s preferentially collimated perpendicular to MSP's spin axis \citep[e.g.,][]{1999ApJ...511..351C,2010MNRAS.404..767C,2010ApJ...715.1282B,2014ApJ...793...97K,2016MNRAS.457.2401C,2017ApJ...842...80K,2018ApJ...857...44K,2018ApJ...855...94P,2019ApJ...874..166C,2023ApJ...954..204K,2024arXiv241202307C}. As a consequence of the spin-up formation process, MSP's spin and orbital momentum vectors are close to alignment. Thus, the companions of MSPs generally experience significant pulsed GeV $\gamma\text{-ray}$ irradiation fluxes.

Nearly 80\% of detected $\gamma\text{-ray}$ MSPs have binary companions \citep{3PC}, the majority of which are He \textcolor{black}{WDs}, but can also be Carbon-Oxygen (CO) WDs, brown dwarfs (BD), main sequence-like stars (MS) or ultralight (UL) companions with mass $\ll0.1M_\odot$ and unknown composition \citep{1992Natur.355..145W,wang2024}. Two interesting classes of binary pulsar systems are black widow (BW) and redback (RB) pulsars \citep{2011AIPC.1357..127R,2013IAUS..291..127R}. These compact binaries typically have orbital periods $<2$ days and are differentiated by the companion masses, $\ll0.1M_\odot$ and $>0.1M_\odot$ for BWs and RBs, respectively. Owing to strong irradiation for many Myr, BW and RB systems have poorly-understood atmospheric structures and composition \citep{2011ApJ...728...95V,2018ApJ...859...54L,2019ApJ...872...42S,2022ApJ...941..199S,2025MNRAS.536.2169S,2025arXiv250511691K}. Their atmospheres are perhaps ablated of lighter elements with possible lithium enrichment from bombardment and spallation by relativistic species or $\gamma\text{-ray}$s \citep{2013ApJ...772L..27M,2015ApJ...804..115R,2022MNRAS.513...71S}. This binary interaction for Gyr timescales thus ought to result in generally more stripped companions than isolated companions of similar mass.

UL companions, which might also exist in triples/multiples \citep{2014Natur.505..520R,2022ApJ...931L...3N}, may form in a high metallicity accretion disk over $>1$ Gyr \citep{Patruno_2017} or via exchange interactions and capture in a cluster \citep{2003Sci...301..193S}. Another possibility for UL companion formation, and particularly for BW systems, is by accretion of an ultra-compact X-ray binary companion's outer layers onto the NS \citep{Bailes_2011}. A severe bias exists against detection of pulsars in ultra-compact binaries (UCBs), owing to Doppler smearing of pulses \citep{1991ApJ...368..504J,2018ApJ...863L..13A,Pol_2021} or infeasible computational cost \citep{2017ApJ...834..106C,2020ApJ...901..156N,2020ApJ...902L..46N}. Studies, accounting for this bias, suggest there may be $\sim7000$ MSP-WD UCBs with $P_{\rm orb} \lesssim 15$ min in the galaxy \citep{Pol_2021}. X-ray studies point to even higher numbers, perhaps $\sim10^5$ UCBs in the bulge \citep{2013A&A...552A..69V}. UCBs also emit gravitational waves (GWs) with frequency $f_{\rm GW} > 10^{-5}$ Hz, and numerous studies estimate that the Laser Interferometer Space Antenna (LISA) will detect $\sim100$ NS-WD UCBs 
\citep{amaroseoane2017,colpi2024,Tauris_2018,Chen_2020,breivik2025,Nelemans_2001,Liu_2014,Breivik_2020}, most of whose pulsars will not be beamed toward Earth. 

\begin{figure*}

\includegraphics[width=0.99\textwidth]{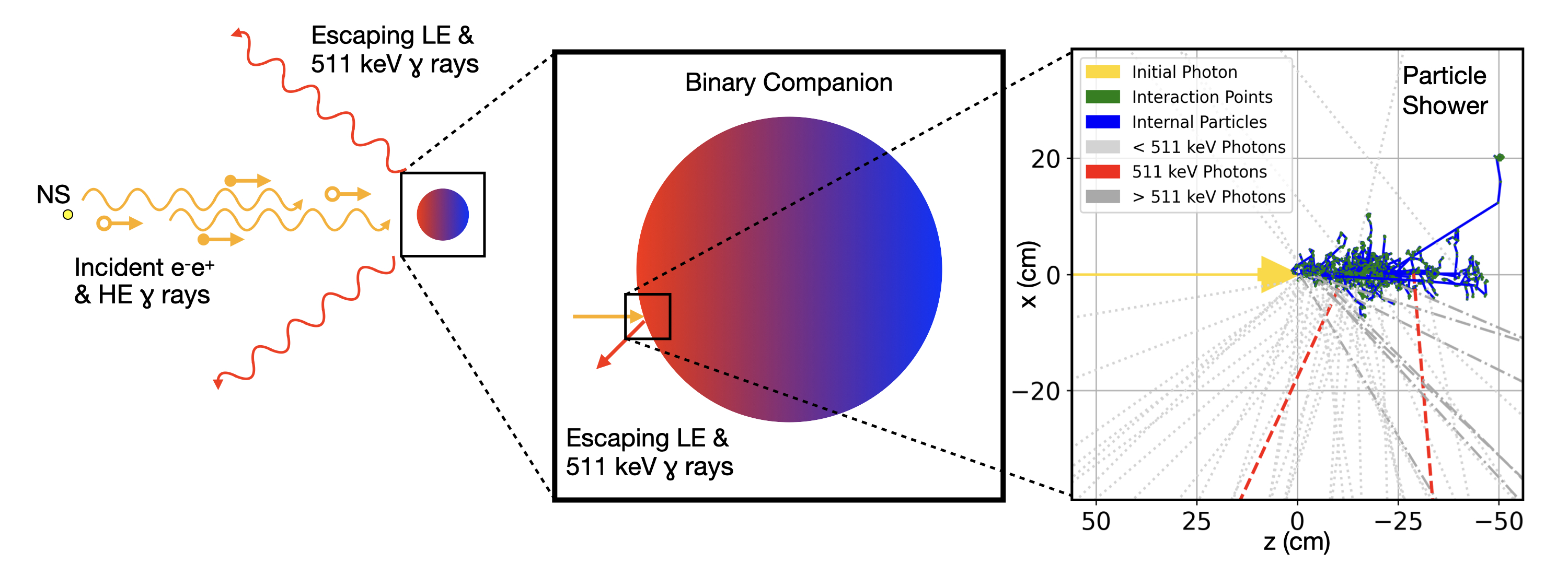}
\caption{Schematic of a MSP irradiating its companion with high energy $\gamma\text{-ray}$s and charged particles. These primary particles bombard the surface of the companion, which are reprocessed into low energy back splash $\gamma\text{-ray}$s, low energy $\rm e^-e^+$ pairs, and 511 keV line emission that escape from the system. In the rightmost panel, $x=0$ corresponds to incident $\gamma\text{-ray}$'s path, and $z=0$ corresponds to the surface of the companion along that path.}
\label{fig:cartoon}
\end{figure*}

In this study, we consider for the first time the bombardment of a range of companion types by the pulsar's primary high energy $\gamma\text{-ray}$s and the resulting shower of secondary lower energy $\gamma\text{-ray}$s and $\rm e^-e^+$ pairs, \textcolor{black}{some of which escape, as depicted in Figure \ref{fig:cartoon}. The secondary particles that escape are called back splash emission.} Prior studies have considered companion heating through bombardment of $\rm e^-e^+$ pairs \citep{Krolik1990}, or bombardment of a companion by TeV protons \citep{Hillas_1984}, but neither conceived of the possibility of 511 keV back splash emission. In Section \ref{sec:methods}, we outline the steps to model 12 representative systems with {\tt Cosima} \citep{MEGAlib}, which is a {\tt Geant4}-based \citep{AGOSTINELLI2003250} Monte Carlo simulator, injecting high energy $\gamma\text{-ray}$s with a fiducial luminosity $L_{\rm MSP}=10^{34}$ ergs/s and a representative spectrum for MSPs in the 3rd {\it Fermi}-LAT Pulsar Catalog \citep{3PC}. In Section \ref{sec:results}, we present the resulting spectra and their geometric, orbital and composition dependencies. Finally, in Section \ref{sec:discussion} we examine the predicted signals from known MSP binary systems.

\section{Methods}\label{sec:methods}

\subsection{Primary Irradiation Channels} \label{sec:channels}

The irradiation and thus the 511 keV line luminosity may be sourced by multiple channels: (i) direct irradiation of the companion's surface by the pulsar’s GeV $\gamma\text{-ray}$s (ii) relativistic $\rm e^-e^+$ pairs produced at polar cap cascades \citep{2011ApJ...743..181H} which might impact the companion (iii) ultra-relativistic 10-100 TeV $\rm e^-e^+$ pairs \citep{2018ApJ...869L..18H,2023NatAs...7.1341H}, which produce the pulsar’s GeV emission via curvature radiation, (iv) $>$TeV ions extracted from the neutron star crust, potentially reaching a few percent of the spin-down luminosity of the MSP \citep{Guepin_2020} (v) intrabinary shock re-acceleration of polar cap pairs to TeV energies, producing  synchrotron and inverse Compton $\gamma\text{-ray}$s \citep[e.g.,][]{2017ApJ...839...80W,2017ApJ...845...42S,2020ApJ...904...91V,2024ApJ...964..109S,2025ApJ...984..146S} which irradiate the companion. Channels (i)-(iv) occur in every $\gamma\text{-ray}$ emitting pulsar in a binary, while channel (v)’s efficacy is largely limited to the spider binaries, the BWs and RBs where shock emission is significant. Thus the total 511 keV line luminosity is \textcolor{black}{$L_{511} = L_{511,i} + L_{511,ii} + L_{511,iii} + L_{511,iv} + L_{511,v}$.}
The maximum 511 keV back splash emission is $L_{511} < \dot{E}\Delta \Omega $, where $\Delta \Omega$ is the solid angle of the companion from the MSP's perspective. Sections \ref{sec:results} and \ref{sec:discussion} focus on $L_{511,i}$ as that contribution's primaries are observationally most well understood and least affected by magnetic fields. \textcolor{black}{Section \ref{sec:L2-5} briefly discusses $L_{511,ii-v}$, but the back splash emission from these components is not straightforward to accurately quantify.} The upshot of our simplified analysis is that our results may be considered conservative. 

\subsection{Companion Size and Composition Specification}\label{Companions}

\textcolor{black}{We survey 12 representative companion models -- 5 planets, 5 main sequence stars, and 2 WDs.} Due to the uncertain composition of the planets in pulsar binary systems, we consider different compositions consisting of He, C, S, Si, and Fe. We assume homogeneous spheres of density 10~g~cm$^{-3}$ and radius $6 \times 10^8$~cm, which \textcolor{black}{yields} approximately Earth's mass and radius. In essence, these models physically represent planets with a dense core and variable crust compositions, but considering a homogeneous sphere is simpler and should not affect the results due to the limited penetration depth into the crust. Figure \ref{fig:cartoon} shows an example photon shower for the C planet, which penetrates $\sim$50 cm below the surface. The penetration depth varies greatly with density, with the planets having their first photon interaction $\lesssim 15$~cm below the surface, compared to $\lesssim 10^8$~cm for the $0.3 M_\odot$ MS model, discussed below. We also neglect atmospheres around the planets, due to stripping of atmospheres from pulsar irradiation leading to negligible interaction probability of gamma rays $> 500$~keV around possible tenuous hot coronae. 

The semimajor axis, $a_{10}$ (units of $10^{10}$ cm), is assumed to be the closest orbit without Roche lobe overflow\footnote{\textcolor{black}{This is a reasonable approximation, because MSPs are spun-up through accretion, so these systems must be not too far from Roche lobe overflow. Indeed, optical studies of BWs and RBs show the companions are close to Roche-lobe filling \citep[e.g. Table 6 in][]{2019ApJ...872...42S}.}}, and is equal to the orbit radius for a circular orbit\footnote{\textcolor{black}{Orbits of compact MSP binaries are low eccentricity owing to tidal interactions and GW radiation \citep[e.g.,][]{1992RSPTA.341...39P}.}}. We use the approximation for Roche lobe radius of \citet{Eggleton}, and we assume NS of mass $1.7~M_\odot$\textcolor{black}{, which is near the median of measured masses of several compact binary MSPs \citep[e.g.,][]{2020mbhe.confE..23L}.} 
The orbital period $P_{\rm orb}$ is based on $M_{\rm NS}, M_{\rm comp}$, and $a_{10}$ as $P_{\rm orb} = 2 \pi \sqrt{R_{\rm orb}^3/\left(G[M_{\rm NS} + M_{\rm comp}]\right)}$, where $R_{\rm orb} = a_{10} \times 10^8$ m, and G is the gravitational constant. 
$a_{10}$ and $P_{\rm orb}$ for each model are listed in Table \ref{tab:fluxes}. 

We consider both C and He WD models. We initially considered a uniform sphere with density $10^6$ g cm$^{-3}$ and radius $7\times10^8$ cm for the C WD, but we suspected that there were truncation errors during the simulation due to the small radiation lengths. Therefore, for the C WD we use the same simulation as the carbon planet and scale the results according to the new mass, companion radius, and $a_{10}$. Assuming C WD fills its Roche lobe led to an orbital period of 39 s, which would be very close to merger, so to have a realistic longer-lived system, we increased $a_{10}$ by a factor of 10. \textcolor{black}{This increases the orbital period to $\approx1200$ s, which is in the range expected for UCBs \citep{Tauris_2018}.}

Additionally, we investigated the effect a dense, presumably gaseous, envelope has on the back splash emission from a C WD companion. The simulated envelope is a single element with density falling exponentially with distance from the C WD surface and exponential scale height of $\sim$ 200 m. In the case of a mostly H layer, the 511 keV emission will be reduced as the incident $\gamma$-rays will predominantly interact with the low-Z envelope (the effect of Z on 511 keV emission is discussed in Section \ref{sec:composition_dependence}). However, it is likely that MSPs harboring WDs will have $Z \gg 1$ layers due to stripping. \textcolor{black}{We found that an evelope with $Z = 6$ causes the back splash emission to vary with respect to the C WD model without such an extended layer by $<0.5\%$. This is unsurprising, because the density at which the mean free path (MFP) of a 1 GeV $\gamma$-ray matches the scale height is $\rho\approx10^{-6}\mathrm{~g~cm^{-3}}$. This occurs within 28 scale heights of the C WD surface, which is less than $0.1\%$ of the C WD radius. Therefore, since the effect of the atmosphere is small, we present a homogeneous C WD with the proviso that the resultant 511 keV emission will differ if the WDs are not stripped or the exponential scale height of the atmosphere is comparable to the companion's radius.} For the He WD, we use the same simulation as the He planet, but we scale the results to a solid angle corresponding to a radius of $2\times10^{10}$ cm and recalculate $a_{10}$. \textcolor{black}{This rescaling is valid in the case where the radiation length is much smaller than companion's radius, because the interaction cross sections for $\gamma$-rays, $e^-$'s and $e^+$'s are independent of density.} The He WD model output caries the same qualification related to stripping.

\begin{figure}[h]

\includegraphics[width=0.95\linewidth]{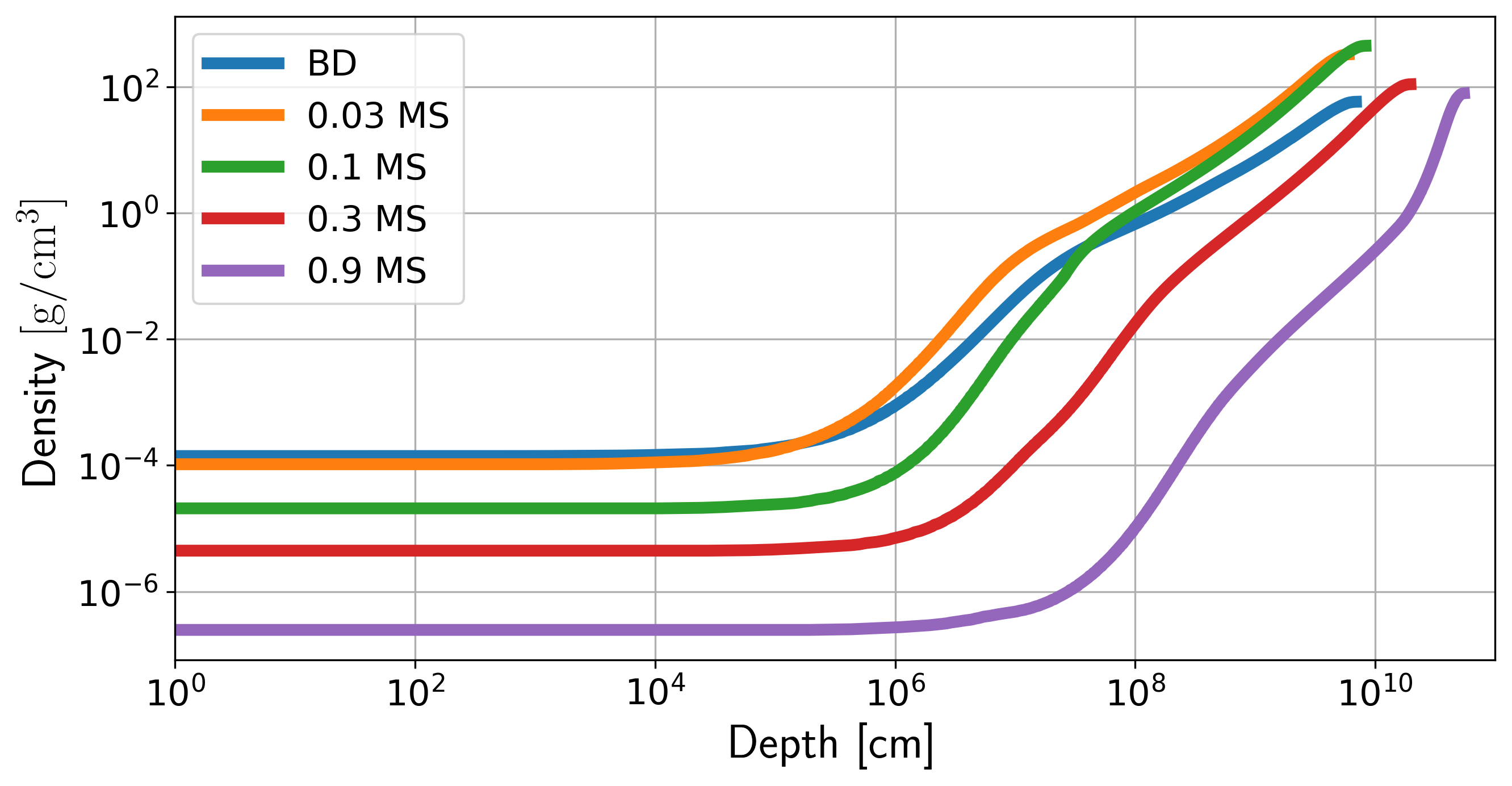}
\caption{Density profiles of BD and MS stars calculated with {\tt MESA}.}
\label{fig:MESA}
\end{figure}

In addition to the homogeneous sphere models, we consider a BD star with mass 0.01 $M_\odot$ and four MS stars with masses of 0.03, 0.1, 0.3, and 0.9 $M_\odot$. Their compositions and densities were determined using the Modules for Experiments in Stellar Astrophysics ({\tt MESA}) software package \citep{Paxton2011,Paxton2013,Paxton2015,Paxton2018,Paxton2019,Jermyn2023}. The model evolution of the 0.1, 0.3 and 0.9 $M_\odot$ MS stars is near ZAMS, while the 0.01 and 0.03 $M_\odot$ star models were evolved for 2 Gyr before stopping\textcolor{black}{, because these models never reach the ZAMS}. The initial conditions for the four MS stars are the standard pre main sequence model from {\tt MESA}, and the intial conditions for the BD star is the standard when creating a generic initial model with the metallicity described in \citet{AG89}. Figure \ref{fig:MESA} shows the density profiles for the BD and MS stars\textcolor{black}{, highlighting the exponential scale heights of each star, and that the densities are roughly constant to depths of $\sim1-100$ km}.

\subsection{Monte Carlo Simulations with {\tt Cosima}}\label{sec:Cosima}

A Monte Carlo simulation for each companion is performed with {\tt Cosima}, the Cosmic Simulator of {\tt MEGAlib}, the Medium Energy Gamma-ray Astronomy library \citep{MEGAlib}, which utilizes the {\tt GEANT4} simulation toolkit \citep{GEANT4} for particle transport. The source for the simulation is assumed to be a far-field point source irradiating a spherical geometry with gamma rays. The spectrum for the gamma rays is a power law with exponential cutoff as in Equation \ref{eqn:dNdE}, where $x = E/E_0$, $E_0 = 2$~GeV, $\Gamma = 1$, $d = 0.6$, $b = 0.9$, and E is the incident gamma ray energy. 

\begin{equation}
    \label{eqn:dNdE}
    \frac{dN}{dE} \propto x^{(-\Gamma + d/b)} \times e^{(d/b^2 + (1-x^b))}
\end{equation}

These values are meant to be representative of the gamma ray spectra in Table 14 of the 3rd Fermi Large Area Telescope Pulsar Catalog (3PC) \citep{3PC}. We assume $L_{\rm MSP}=10^{34}$ erg/s as a representative value, which may be readily scaled up or down. 

MSPs have pulsed GeV gamma ray emission with $P_{\rm MSP} < 30$ ms, but the pulsed emission is treated as continuous in the simulations due to smeared arrival times at the companions. At minimum, photon arrival and escape times for one of the planets with $R_{\rm comp} = 6\times10^8$ cm are smeared by up to 40 ms by the showers, \textcolor{black}{because of light travel time delays associated with the (unresolved) star's curved surface, $\Delta T = 2R_{\rm comp}/c$. This suppresses} periodic fluctuations of the back splash emission modulated with the spin of the MSP.

\begin{figure}

\includegraphics[width=0.95\linewidth]{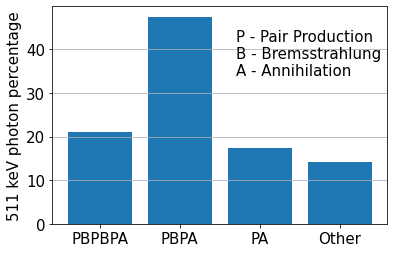}
\caption{Histogram of the dominant pathways for escaped 511 keV photons for a representative C WD. These represent multiple generations of $e^-e^+$ pair production and $\gamma$-ray production through bremsstrahlung that concludes with $e^-e^+$ annihilation, creating an escaping 511 keV $\gamma$-ray.}
\label{fig:Generations_Bar}
\end{figure}

\begin{figure*}[htbp]

\includegraphics[width=0.95\linewidth]{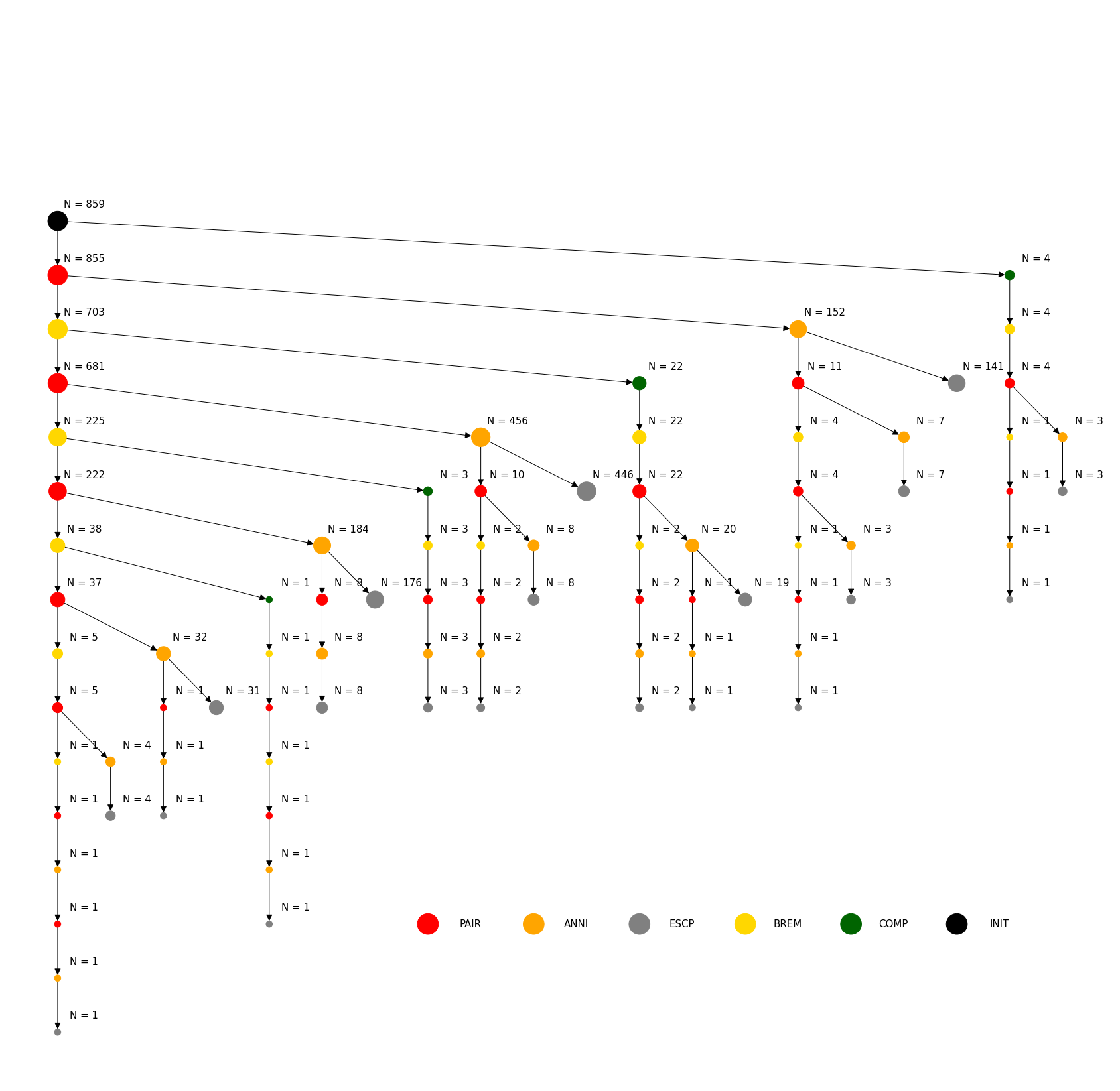}
\caption{This figure charts the genealogy and evolution of 10,000 initial photons that result in escaping 511 keV photons for the C planet system. Each row is a generation and each branch represents a different path. The black vertex is the initial photon (INIT), green vertices are photons producing Compton scattered electrons (COMP), yellow vertices are electrons or positrons producing Bremsstrahlung photons (BREM), orange vertices are electron-positron annihilation to produce photons (ANNI), red vertices are photons producing electron-positron pairs (PAIR), and grey vertices are escaping 511 keV photons.}
\label{fig:generations}
\end{figure*}

Each primary $\gamma\text{-ray}$ that interacts with the companion produces a shower of daughter $\gamma\text{-ray}$s, electrons, and positrons that are all tracked until absorption or escape in 3D. The genealogy of the most prominent interactions that yield 511 keV photons for a representative C WD model are shown in Figure \ref{fig:Generations_Bar}. Most 511 keV production stems from the 1st, 2nd, or 3rd generation of pair production and bremsstrahlung, terminating with $\rm e^-e^+$ annihilation. This \textcolor{black}{suggests the dominance continuous Coulomb (not depicted) and bremsstrahlung losses of charges and the absence of scattering interactions by parent photons of the 511 keV $\gamma$-rays.} The full range of pathways is detailed in Figure \ref{fig:generations}. 

\subsubsection{Unimportance of Plasma Effects} \label{sec:plasma}

{\tt Geant4}, and by extension {\tt Cosima}, assumes non-ionized material rather than a plasma for its test particle shower simulation. The stopping power of ionized gas or coronae around companions in pulsars to $\gamma\text{-ray}$s is generally negligible. The primary $\gamma\text{-ray}$s and associated particle energies are also well above where most atomic transitions are significant. Thus, to a good approximation, scattering of photons by electrons in the simulation behaves as that by free electrons, and Coulomb losses for created pairs are the dominant loss mechanism in the incipient shower\footnote{The {\tt Cosima} simulations of tagged ``ionization" losses in a nonionized material are compatible with the Coulomb stopping power of a cold electron-ion plasma.}. Collective effects here may be neglected in the limit that the Debye length, $\lambda_D = \sqrt{\epsilon_0 k_B T_e/(n_e q_e^2)}$, is shorter than the mean free path ($\epsilon_0$ is the permittivity of free space, $k_B$ the Boltzmann constant, $T_e$ the electron temperature, $n_e$  the electron number density, and $q_e$ the electron charge). Charged particle energies are generally well above typical background plasma thermal energies, $kT_e \ll 1$~keV. As realized temperatures are not relativistic, thermal line broadening will also be minimal. The mean free path is approximated by the continuous slowing down approximation (CSDA) range, $R_{\rm CSDA}$ used by NIST ESTAR~\citep{ESTAR} as  $\lambda_{\rm MFP} = R_{\rm CSDA}/\rho$, where $\rho = m_p A n_e/Z$ is the material density, $m_p$ is the proton mass and $A$ is the atomic mass number of the material,
\begin{equation}
    \label{eqn:MFP Ineq}
    \begin{split}
    \frac{\lambda_D}{\lambda_{\rm MFP}} &= \frac{\sqrt{\epsilon_0 k_B T_e m_p A \rho}}{q_e R_{\rm CSDA}\sqrt{Z}} \\ &= 2.8\times10^{-5} ~T_{6,e}^{1/2} ~\rho_{1}^{1/2} ~R_{-3,\mathrm{CSDA}}^{-1}~A^{1/2}~Z^{-1/2} \\&\ll 1,
    \end{split}
\end{equation}

\textcolor{black}{where $T_{6,e} = T/10^6 ~\mathrm{K}$, $\rho_1 = \rho/10\mathrm{~g~cm^{-3}}$, and $R_{-3,\mathrm{CSDA}} = R_\mathrm{CSDA}/10^{-3} \mathrm{~g~cm^{-2}}$.}
The condition Equation \ref{eqn:MFP Ineq} is generally satisfied for BD, MS, and WD models, including their ionized atmospheres. The planets ought not be appreciably ionized, and their atmospheres should be both lower density and lower temperature than the extremes of the stellar companion models. For the BD and MS models, the highest product of density and temperature is in the core of the 0.1 $M_\odot$ MS star at $~450 \rm ~g/cm^3$ and $~2\times10^7$~K. \textcolor{black}{The C WD of $10^6$~K surface temperature is a fiducial upper bound to the temperature of irradiated WDs. The influence of} inner structure is negligible due to the high density and limited penetration depth. The smallest $R_{\rm CSDA}$ are thus realized for low energy electron/positrons that could be produced in the shower of the primary high energy $\gamma\text{-ray}$s.

\subsubsection{Unimportance of Magnetic Field Effects}\label{sec:B}

\textcolor{black}{Magnetic fields are neglected; however as we show it is likely that surface magnetic fields have a negligible influence on the $\gamma$-ray showers. A relativistic electron of Lorentz factor $\gamma$ in a magnetic field $B$ loses energy by magnetic bremsstrahlung (synchrotron) over a length scale $\lambda_\text{B} = c\gamma/\dot{\gamma}$, where $c$ is the speed of light. Here $\dot{\gamma} \sim 4\sigma_\text{T} c U_\text{B} \gamma^2/(3 m_e c^2)$ where $\sigma_\text{T}$ is the Thomson cross section, $m_e$ is the mass of an electron, $U_\text{B}=B^2/(8\pi\mu_0)$ and $\mu_0$ is the permeability of free space.  We find $\lambda_\mathrm{MFP}/\lambda_B \ll 1$ as in Equation \ref{eqn:lambda_B} even for extreme values where $B_6 = B/10^6\mathrm{~G}$ and $\gamma_2 = \gamma/100$.}


\begin{equation}\label{eqn:lambda_B}
    \begin{split}
        \frac{\lambda_\mathrm{MFP}}{\lambda_B} &= \frac{R_\mathrm{CSDA}\sigma_T B^2 \gamma}{6\pi\mu_0m_ec^2\rho} \\
        &=3.4\times10^{-11} ~R_{-3,\mathrm{CSDA}}~B_6^2~\gamma_2~\rho_1^{-1} \ll 1
    \end{split}
\end{equation}

Thus, a 100 MeV positron in a $10^6$ G magnetic field has $\lambda_\text{B}~=~1.5\times10^6$~cm, which is much larger than $\lambda_\text{MFP}~=~4$~cm in $10 \mathrm{~g/cm^3}$ of C. Magnetic bremsstrahlung may be an important effect when considering charged particle showers, but that is not the focus of this work.

\section{Results}\label{sec:results}

We define the 511 keV production efficiency, $\epsilon$, in Equation \ref{eqn:efficiency}, where $\dot{N}_{511}$ is the total luminosity of 511 keV photons escaping the system in units of $\rm photons/s$, and $\int_{E_{\rm min}}^{E_{\rm max}} \frac{dN}{dE} dE = 1$. Table \ref{tab:fluxes} lists $\epsilon$ for each model. Equation \ref{eqn:efficiency} also shows that, keeping the companion and MSP properties constant, $\dot{N}_{511}$ is linear with $\Delta\Omega$, so there will be more back splash 511 keV emission from a compact binary or a binary with a larger companion.

\begin{equation}
    \epsilon = \dot{N}_{511} \times \frac{\int_{E_{\rm min}}^{E_{\rm max}}\frac{dN}{dE}E dE}{L_{\rm MSP}} \times \frac{4\pi}{\Delta \Omega}
    \label{eqn:efficiency}
\end{equation}

\begin{table}
\scriptsize
\begin{ruledtabular}
\begin{tabular}{|c|c|c|c|c|c|c|}
\hline
Model     & $a_{10}$ & $P_{\rm orb}$ & $R_{\rm comp}$ & $\epsilon$ (\%) & Peak $\dot{n}_{511}$ & $\dot{N}_{511}$\\\hline
Units     & --       & $10^4 \rm s$  & $10^8 \rm cm$  &  \%             &${\rm ph/cm^2/s}$     & ${\rm ph/s}$ \\\hline
He planet & 8.8      & 1.1           & 6              & 2               & $10^{-12.0}$         & $10^{29.8}$   \\\hline
C planet  & 8.8      & 1.1           & 6              & 10              & $10^{-11.4}$         & $10^{30.5}$   \\\hline
Si planet & 8.8      & 1.1           & 6              & 42              & $10^{-10.8}$         & $10^{31.2}$   \\\hline
S planet  & 8.8      & 1.1           & 6              & 54              & $10^{-10.6}$         & $10^{31.3}$   \\\hline
Fe planet & 8.8      & 1.1           & 6              & 126             & $10^{-10.3}$         & $10^{31.6}$   \\\hline
He WD     & 9.7      & 1.2           & 200            & 2               & $10^{-9.1}$          & $10^{32.8}$   \\\hline
C WD      & 2.3      & 0.12          & 7              & 10              & $10^{-10.1}$         & $10^{31.8}$   \\\hline
0.01 $M_\odot$ BD & 8.0 & 0.94       & 69             & 1               & $10^{-10.0}$         & $10^{31.8}$   \\\hline
0.03 $M_\odot$ MS & 4.9 & 0.45       & 60             & 1               & $10^{-9.7}$          & $10^{32.2}$   \\\hline
0.1 $M_\odot$ MS  & 4.7 & 0.41       & 83             & 1               & $10^{-9.5}$          & $10^{32.4}$   \\\hline
0.3 $M_\odot$ MS  & 8.2 & 0.90       & 200            & 1               & $10^{-9.2}$          & $10^{32.8}$   \\\hline
0.9 $M_\odot$ MS  & 17  & 2.4        & 550            & 1               & $10^{-8.9}$          & $10^{33.0}$   \\\hline

\end{tabular}
\end{ruledtabular}
\caption{\label{tab:fluxes} For each companion model, the number of photons between 505 and 515 keV per interacting primary $\gamma\text{-ray}$ is reported. The 511 line flux assumes a pulsar luminosity of $10^{34}$ erg/s, and a distance of 100 pc from Earth.}
\end{table}

The annihilation line $\gamma\text{-ray}$ emission is not isotropic, so in Equation \ref{eqn:diff flux} we define a differential $\gamma\text{-ray}$ emissivity, $\dot{n}_{511}(\phi,i)$ (units of $\rm photons/s/sr$), escaping at an azimuthal angle in the orbital plane, $\phi$, and inclination angle, $i$. $\phi = 0$ is defined as directly towards the MSP, and $i = 0$ is orthogonal to the orbital plane. Table \ref{tab:fluxes} lists $\dot{N}_{511}$ and the maximum of $\dot{n}_{511}(\phi,i = \pi/2)$ for each model, assuming a distance of 100~pc from Earth. 

\begin{equation}
    \label{eqn:diff flux}
    \dot{N}_{511} = \iint d\phi \, di \sin(i) \dot{n}_{511}(\phi,i)
\end{equation}

\begin{figure}

\includegraphics[width=0.95\linewidth]{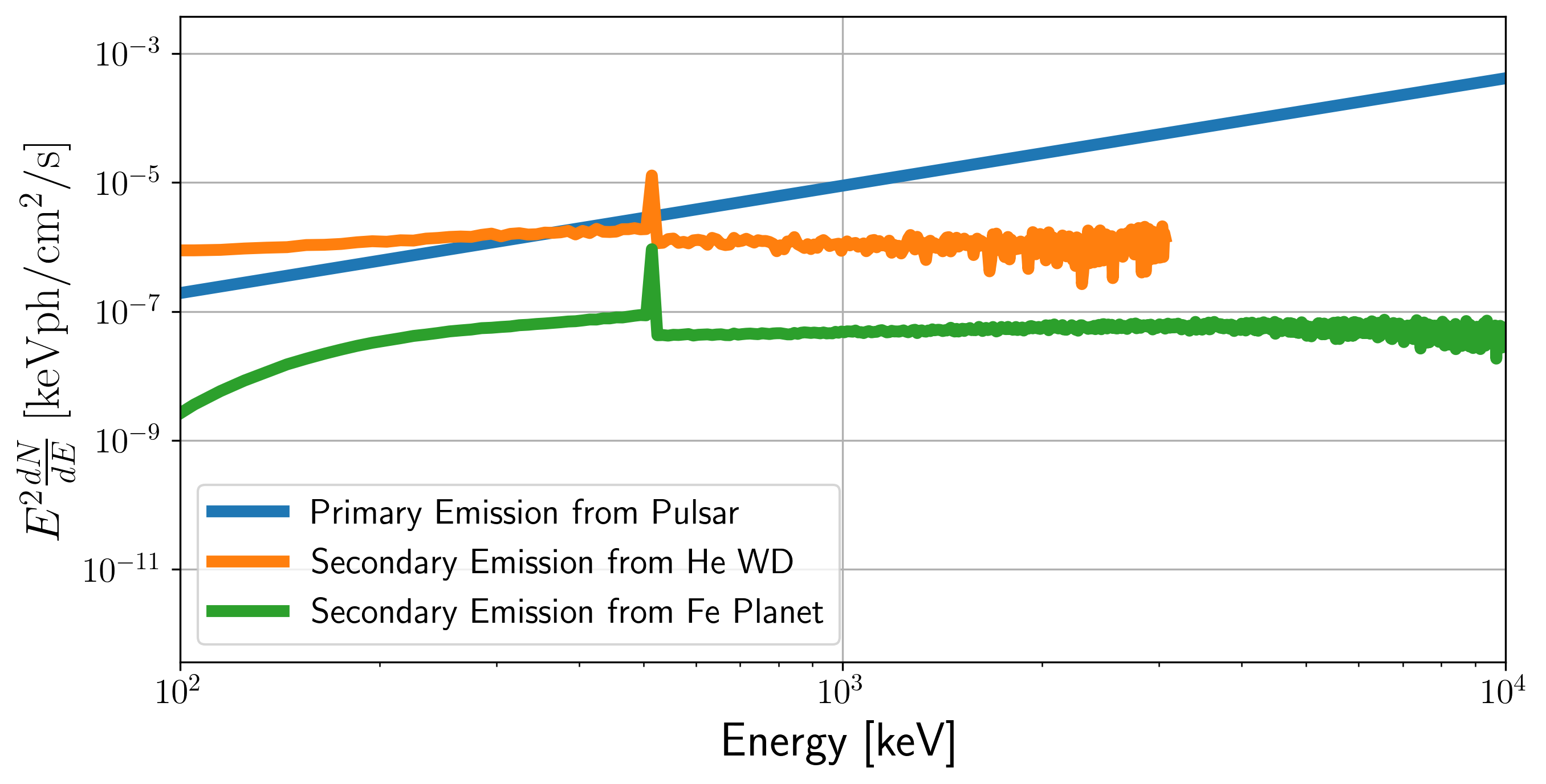}
\caption{\textcolor{black}{Primary incident (blue) and secondary emission for an irradiated He WD (orange) and Fe planet (green).} The MSP spectrum is defined by Equation \ref{eqn:dNdE} with $x = E/E_0$, $E_0 = 2$~GeV, $\Gamma = 1$, $d = 0.6$, $b = 0.9$, and E is the incident $\gamma\text{-ray}$ energy. This system is assumed to be 100 pc from Earth, and the pulsar luminosity is $L_{\rm MSP} = 10^{34}$~ergs/s. The normalization for this spectrum is averaged over all emission directions, and the instantaneous spectrum can vary by viewing angle.}
\label{fig:Combined Spectrum}
\end{figure}

\begin{figure*}

\includegraphics[width=0.95\linewidth]{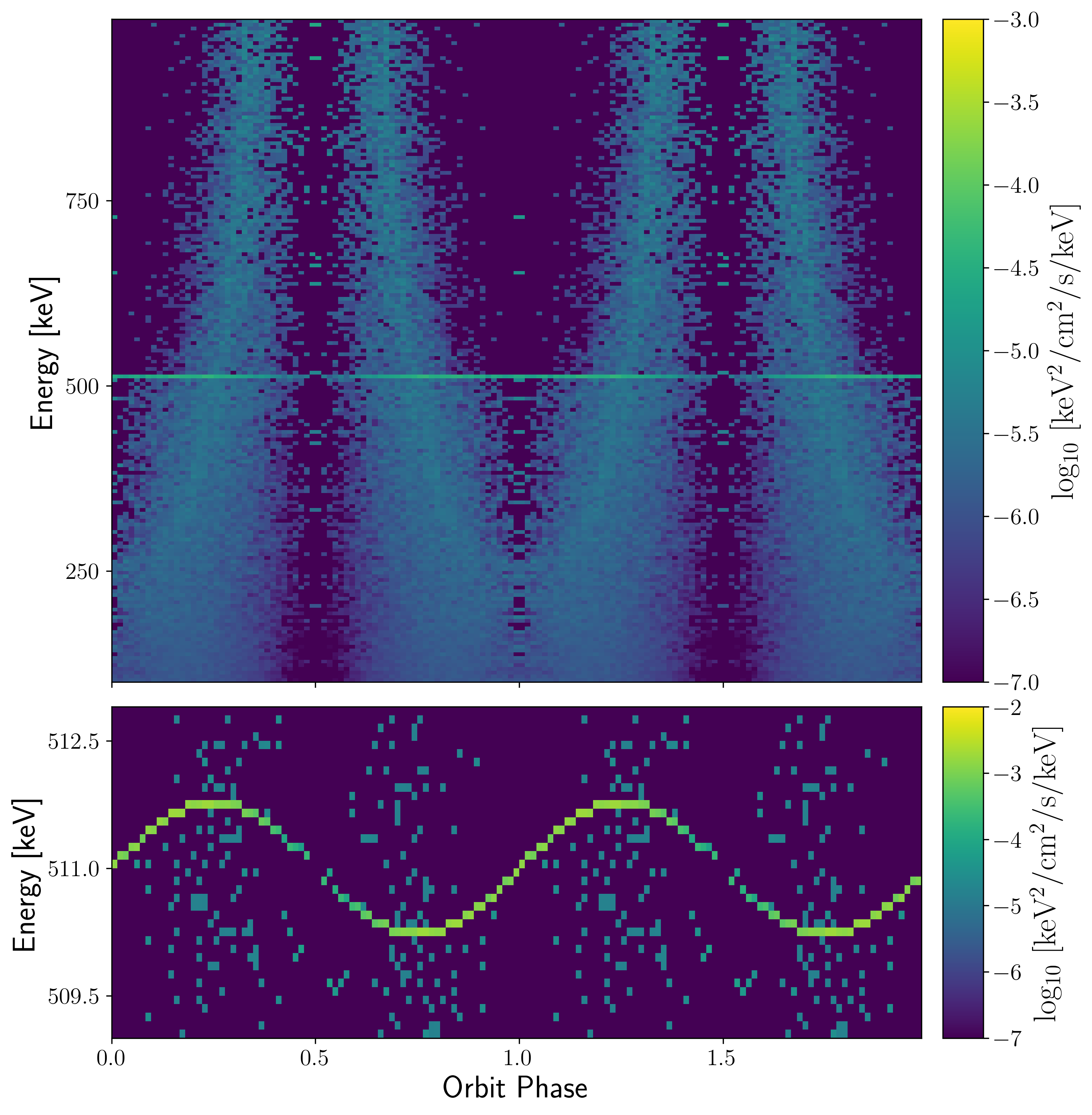}
\caption{The back splash spectrum, and radial Doppler shifts (for zero systemic/peculiar velocity), for a He WD-MSP binary 100 pc from Earth viewed edge-on over the course of an orbit. Orbit phase = 0 corresponds to the inferior conjunction of the MSP (the MSP between the observer and the companion). The 511 keV line shifts by $\sim1~\rm keV$, which could substantially increase the sensitivity of coherent searches for the line emission if the Doppler shift can be resolved.}
\label{fig:2D Spectrum}
\end{figure*}

\begin{figure}[]
    \centering
    \subfloat[]{\includegraphics[height=0.3\textwidth]{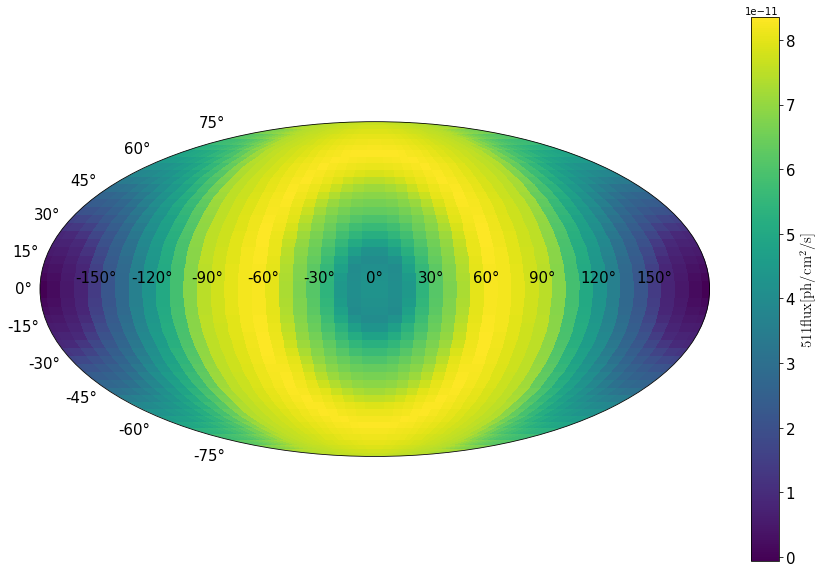}\label{fig:eye}}\hskip1ex
    \subfloat[]{\includegraphics[height=0.3\textwidth]{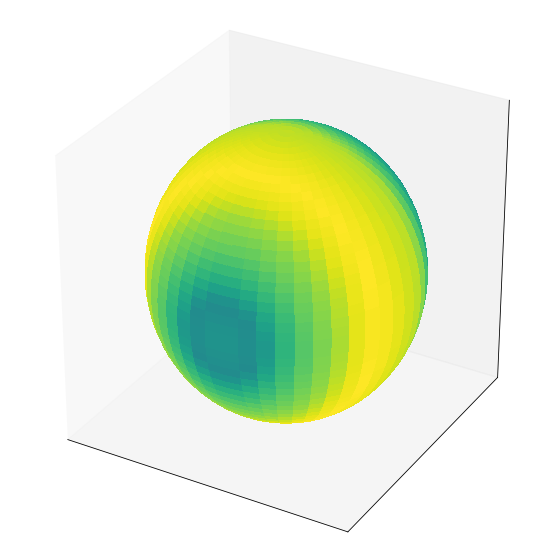}\label{fig:sphere}}
    \caption{\textcolor{black}{The sky distribution $\dot{n}_{511}(\phi,i)$ for the C WD model (placed at 100 pc). The center of the ``eye" in \textbf{(a)} is the inferior conjunction of the MSP (the MSP between the observer and the companion). \textbf{(b)} shows the same fluxes as \textbf{(a)}, projected onto the surface of a sphere, where a brighter spot means there is more 511 keV emission in that direction. The relative difference in flux of the center of the ''eye" and bright ring correlates with the atomic number, Z, of the companion. Thus observations could constrain the companion composition. This effect is discussed in more detail in Section \ref{sec:composition_dependence}.}}
\end{figure}

The He WD and massive star systems have the brightest 511 keV line at $10^{-8.9}$ and $10^{-9.1} \rm ~ph/cm^2/s$ at 100~pc, respectively. \textcolor{black}{A spectrum for the secondary emission from the He WD system is shown alongside the initial MSP spectrum and that of a Fe planet in Figure \ref{fig:Combined Spectrum}, demonstrating that the back splash emission rises above the MSP's primary continuum emission (which are generally not beamed toward the observer in similar directions) by a factor of $\sim10$ at 511~keV for the He WD. For fainter models like the Fe planet,} the line may not rise above the pulsar emission, complicating searches for the emission from these systems, and indicating that searches would be more sensitive to binary MSP systems with harder primary $\gamma\text{-ray}$ spectra than our assumption or whose pulsar emission is not beamed toward Earth. Figure~\ref{fig:2D Spectrum} shows the secondary emission spectrum for the same system as a function of orbit phase. An orbital Doppler shift modulates the line by $\lesssim1$ keV, neglecting the finite size of the companion. \textcolor{black}{Figure \ref{fig:eye}--\ref{fig:sphere} shows the sky distribution $\dot{n}(\phi,i)$ for the C WD model, where the center of the circle corresponds to emission directly towards the MSP.
From the distribution $\dot{n}(\phi,i)$, it is established that 511 keV line flux and it orbital phase dependence also is influenced by an observer's orbital inclination angle.}

\subsection{Appraisal of Other Irradiation Channels}\label{sec:L2-5}

We briefly consider $L_{511,ii-v}$ and address why they are not included in the main analysis. 

The second channel, $L_{511,ii}$, is the effect of relativistic $e^+e^-$ pairs produced at the polar cap in the MSP and advected out into the wind. To model this, we replaced the gamma ray flux from the simulation with incident $e^+e^-$ pairs with a luminosity, $L_{e^+e^-, \rm pc} = 10^{34} \rm \,erg/s$, and a spectrum that follows that of a MSP with $P_{\rm rot} = 2 \rm ~ms$, 0 degree offset, and $B = 2 \times 10^9 \rm ~G$ from \citet{Harding_2011}. For this simulation, we injected 10,000 $e^-e^+$ pairs with the Fe planet model and tracked the escaping gamma rays.

The third channel, $L_{511,iii}$, is the effect of ultra-relativistic 10-100 TeV $e^-e^+$ pairs \citep[e.g.,][]{2021JCAP...02..030L,2024ApJ...964..109S}. Due to limitations with the simulation, simulating a statistically significant number of 100 TeV particles would take an unreasonable amount of time. Therefore, we instead simulated 10,000 pairs each of monochromatic 1, 10, and 100 GeV particles with the Fe planet model with $L_{e^+e^-, \rm TeV} = 10^{34} \rm erg/s$. Then, we project $L_{511}$ from these lower energy particles to 10 TeV to get a rough estimate of $L_{511,iii}$.

The fourth channel, $L_{511,iv}$, is the effect of protons accelerated to GeV energies. Our simulations neglected nuclear effects, including activation of the companion crust by charged particle bombardment. As such, it would have been unreasonable to include proton bombardment without nuclear interactions. A future study will reconsider the impact of nuclear interactions on the 511 keV emission as well as the presence of other activation lines.

The fifth channel, $L_{511,v}$, is the result of shock formation between the MSP and the companion for BW and RB systems. The shock produces a power law with exponential cutoff spectrum with unknown cutoff energy between $\gtrsim 100$ keV and $\sim 150$ MeV \citep[e.g.,][]{2020ApJ...904...91V,2022ApJ...933..140C,2024MNRAS.534.2551C,2024ApJ...964..109S}. For this simulation, we produce 10,000 photons on half of the sky from the perspective of the companion as opposed to a far field point source. The spectrum used is the same as in Equation \ref{eqn:dNdE} with $E_0$ = 150 MeV, $\Gamma$ = 1.67, d = 0.1, and b = 3. This simulation assumes a gamma ray luminosity of $L_{\rm shock} =  10^{32} \rm erg/s$, and uses the Fe planet model. This assumption represents the upper limit of $L_{511,iv}$, because a lower cutoff energy potentially drastically reduces the line production efficiency.

$L_{511}$ is presented in Table \ref{tab:Channels} for all of the considered production channels along with the steps used to reach $L_{511,iii}$. Only $L_{511,i}$ is presented in the other sections, because $L_{511,ii}$ is negligible, $L_{511,iii}$ requires projecting the 511 efficiency over two orders of magnitude, this analysis neglected nuclear interactions for $L_{511,iv}$, and the cutoff energy of the primary gamma ray spectrum for $L_{511,iv}$ is unconstrained by observations. Taking each of these simulations at face value suggests that $L_{511}$ could be substantially larger than $L_{511,i}$, but the uncertainties on companion composition and size represent larger uncertainties than those from neglecting $L_{511,ii,iii,iv,v}$.

\begin{table}[h]
\begin{ruledtabular}
\begin{tabular}{|c|c|}
\hline
Channel& $L_{511} {\rm [erg/s]}$\\\hline
Primary GeV $\gamma$-rays, $L_{511,i}$ & $10^{25.5} L_{\rm MSP,34}$\\\hline
*Primary $\rm e^- e^+ Pairs$, $L_{511,ii}$ & $10^{23.4} L_{\rm ee, 34}$\\\hline
Mono. 1 GeV $\rm e^+ e^- Pairs$ & $10^{22.5} L_{\rm 34}$\\\hline
Mono. 10 GeV $\rm e^+ e^- Pairs$ & $10^{23.2} L_{\rm 34}$\\\hline
Mono. 100 GeV $\rm e^+ e^- Pairs$ & $10^{23.9} L_{\rm 34}$\\\hline
Mono. 1 TeV $\rm e^+ e^- Pairs$ (proj.) & $10^{24.6} L_{\rm 34}$\\\hline
Mono. 10 TeV $\rm e^+ e^- Pairs$ (proj.), $L_{511,iii}$ & $10^{25.3} L_{\rm 34}$\\\hline
**Shock MeV $\gamma$-rays, $L_{511,v}$ & $10^{25.7} L_{\rm Shock,32}$\\\hline
\hline

\end{tabular}
\end{ruledtabular}
\caption{\label{tab:Channels} For the ideal Fe planet model, we calculate $L_{511}$ for each of the primary radiation channels. * Solid Blue line Fig 10 of \citet{Harding_2011}. ** Power law with exponential cutoff as in Equation \ref{eqn:dNdE} with $E_0 = 150$ MeV, $\Gamma = 1.67$, $d = 0.1$, and $b = 3$.}
\end{table}

\subsection{Composition Dependence of Pulse Shape} \label{sec:composition_dependence}

\begin{figure}
\includegraphics[width=0.95\linewidth]{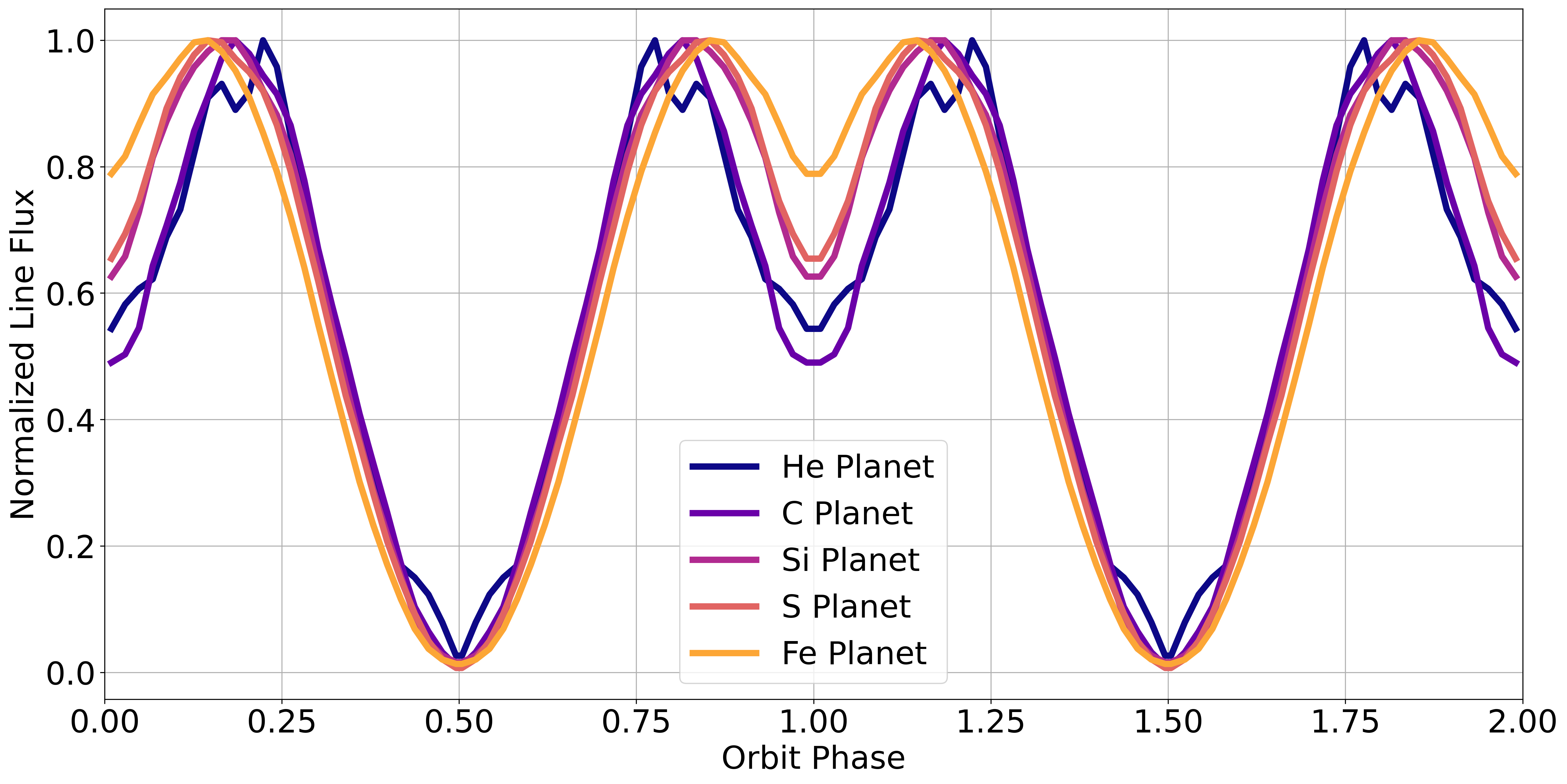}
\caption{The normalized 511 keV flux as a function of orbit for 5 planet models at $i=90^\circ$.}
\label{fig:Flux_by_orbit_normalized}
\end{figure}

\begin{figure}[h]

\includegraphics[width=0.95\linewidth]{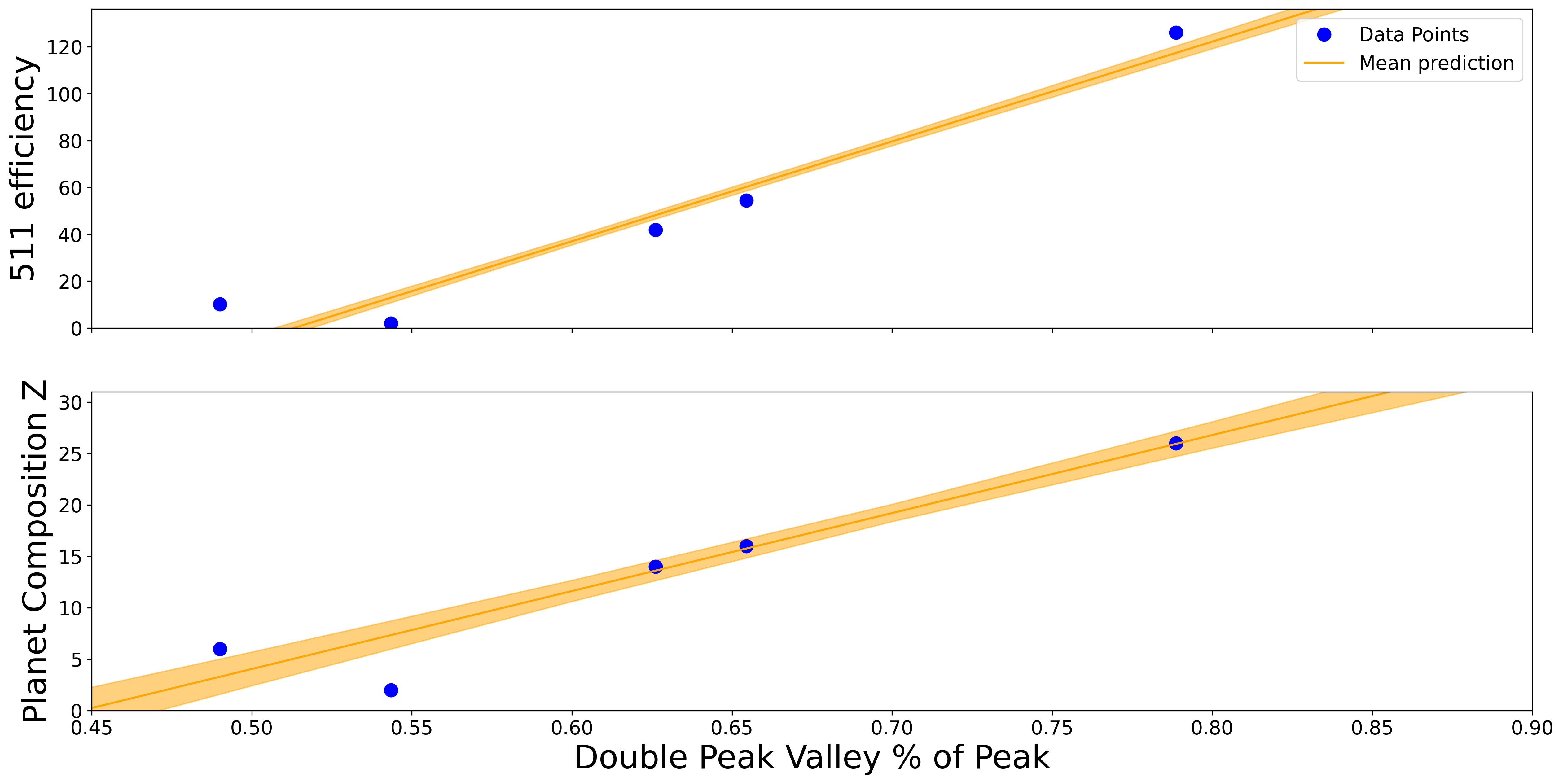}
\caption{$\epsilon$ is shown as a function of of $\frac{\dot{n}_{511}(0,\frac{\pi}{2})}{\max(\dot{n}_{511}(\phi,\frac{\pi}{2}))}$ for each planetary model. All five models have the same density and radius with the only difference being the composition of the companion.}
\label{fig:efficiency_dependence}
\end{figure}

The orbital phase dependence of the 511 keV emission correlates with the atomic number, Z, of the planetary models. Figure \ref{fig:Flux_by_orbit_normalized} shows $\dot{n}_{511}(\phi,i = \pi/2)$, normalized such that the peak is set to 1, and only for the fluxes from the five planet models. The pulse shapes feature a double peak near when the MSP eclipses the planet. \textcolor{black}{The shapes also demonstrate that the relative depth of the double peak with respect to the overall maximum flux decreases with increasing Z.} Figure \ref{fig:efficiency_dependence} shows that the depth of the double peak also correlates with the 511~keV production efficiency, $\epsilon$, which is expected given that Table \ref{tab:fluxes} shows $\epsilon$ increasing with Z. 

The origin of the double peak feature comes from the fact that Coulomb losses, Bremsstrahlung, and photon scattering cross sections all scale with Z, but pair production scales roughly with $\rm Z^2$. For higher Z materials, the electron positron pairs and subsequently the 511 keV photons are produced within fewer radiation lengths of the companions surface for higher Z materials. This explains the overall increase in $\epsilon$ with Z. Coupled with the fact that the incident primary photons are traveling parallel, it explains the double peak and its Z dependence. Let the MSP be at z = $+\infty$, then the particle showers are all going to be in the --z direction along the path of the primary photon. For a photon that interacts near the center of the face of the companion, the shower is furthest from the surface, but for a photon that interacts near the edge of the face of the companion, the shower is closer to the surface, because the surface curves to be parallel with the z axis at the edge of the face of the companion. Since the 511 keV photons are produced within fewer radiation lengths of the surface for high Z materials, the effect of the curvature of the face of the companion is reduced, and this leads to a shallower double peak than for low Z companions.

\section{Discussion}\label{sec:discussion}

\subsection{Observability of Known Systems} \label{sec:observability}

\begin{figure*}
    \centering
    \includegraphics[width=1\linewidth]{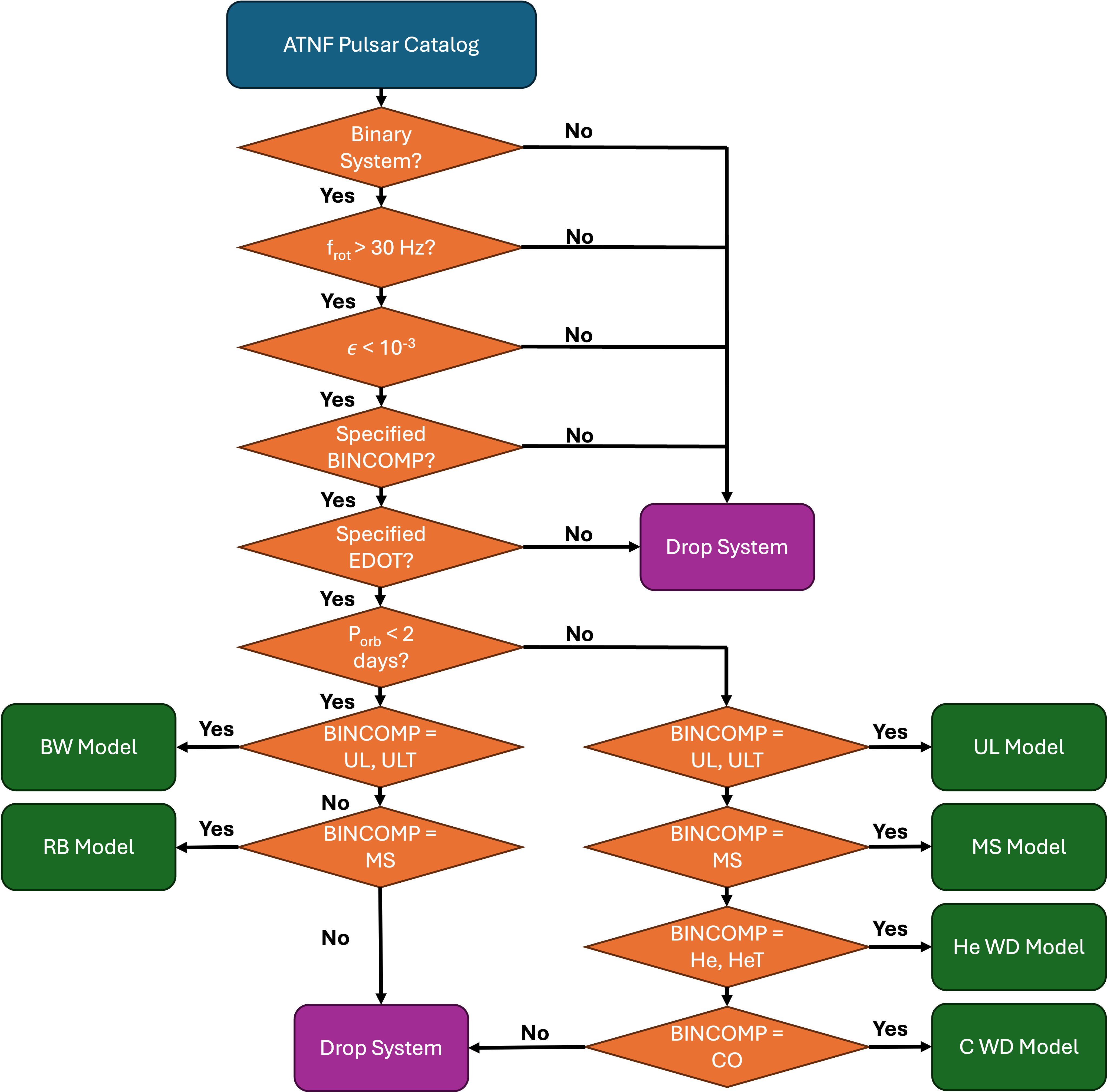}
    \caption{\textcolor{black}{Procedure for filtering for known systems and estimated 511 keV fluxes. Binary MSP systems were selected from those listed in the ATNF Pulsar Catalogue \citep{Manchester_2005} and its entry definitions. The systems are divided based on the `BINCOMP' column. Systems defined as `He' or `HeT' use the He WD model, those defined as `CO' use the C WD model, and those defined as `MS' happened to be RB systems with $ P_{\rm orb}< 2$ days and used the MS model with the closest mass. The systems defined as `UL' or `ULT' were further divided into BWs, if $P_{\rm orb} < 2$ days, or UL otherwise.}}
    \label{fig:ATNF_Flowchart}
\end{figure*}

We now discuss the consequences and implications of the results above by considering binary MSPs low eccentricity in the ATNF Pulsar Catalogue \citep{Manchester_2005}. \textcolor{black}{Figure \ref{fig:ATNF_Flowchart} shows the decision tree to select and categorize the systems to fit the BW, RB, UL, MS, He WD, and C WD models.} We assume 1.7~$M_\odot$ NSs and that the stellar companions' radii are the value listed in Table \ref{tab:fluxes}. The BW and UL systems are treated differently, \textcolor{black}{because their compositions and densities are unknown. To account for a range of possibilities, we} make three different assumptions for the BW and UL companions: a) the companion density is $10 \rm \, g/cm^3$, b) the companion density is $30 \rm \, g/cm^3$, and c) the companion radius is $6\times10^8 \, \rm cm$. For reference, the companion orbiting J1719--1438 has a minimum density of 22 g/cm$^3$~\citep[a `diamond' planet,][]{Bailes_2011}. In any case where the Roche lobe radius is less than the assumed radius, we use the Roche lobe radius instead. We assume $L_{\rm MSP} = 0.2 \dot{\rm E}$ and the same primary $\gamma\text{-ray}$ spectrum as in the previous section. The systems and their parameters are listed in the Appendix. 

\begin{figure*}
\centering
\includegraphics[width=0.8\textwidth]{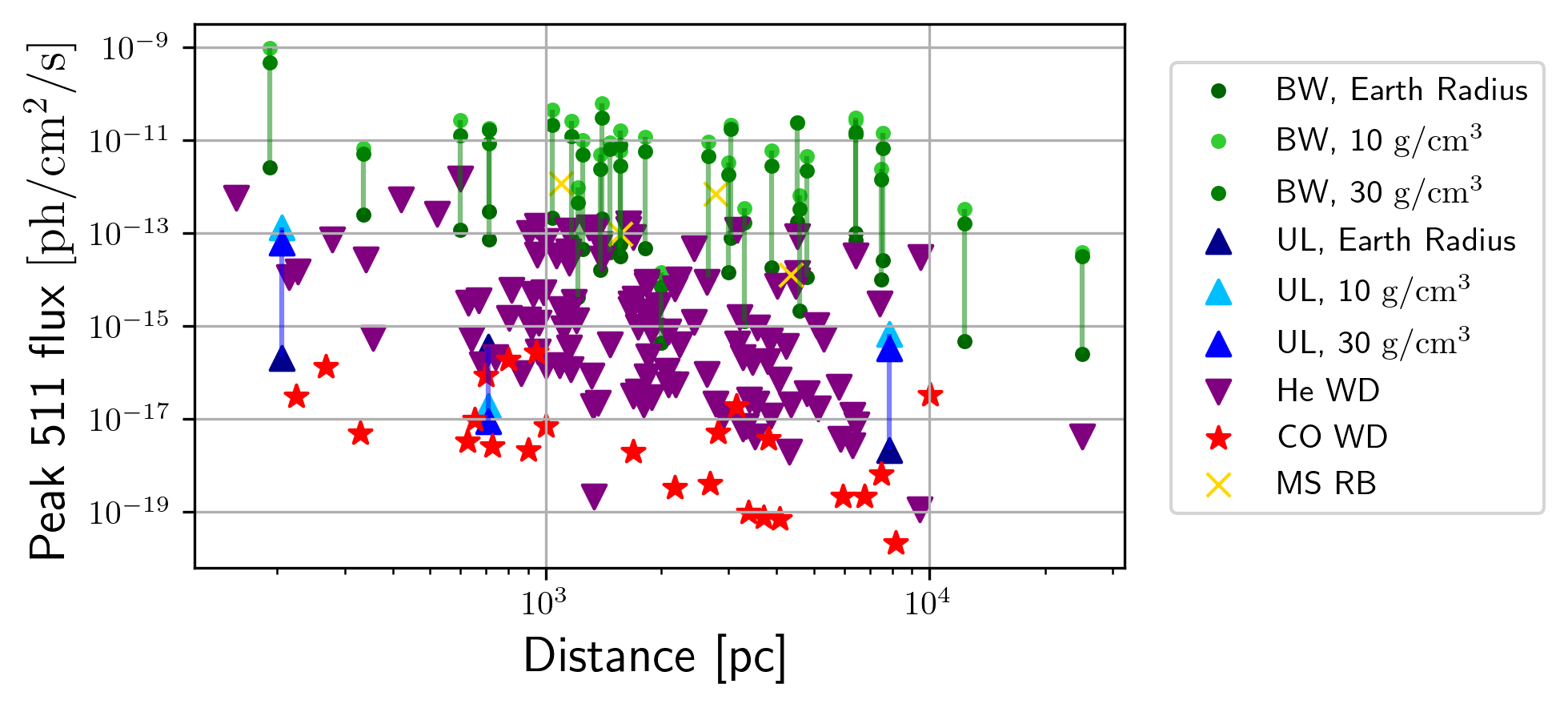}
\caption{The estimated flux for the MSP binary system in the ATNF Pulsar Catalogue. The BW and UL systems are modeled using three size assumptions\textcolor{black}{, because their true density is unknown.} a) The companion radius is $6\times10^8 \rm ~cm$. b) The companion density is $10 \rm ~g/cm^3$. c) The companion density is $30 \rm ~g/cm^3$. In any case where the Roche lobe radius is less than the assumed radius, we use the Roche lobe radius instead. The BW systems are the most promising known candidates for observation.}
\label{fig:Real_Systems}
\end{figure*}

The maximum of $\dot{n}_{511}(\phi,i = \pi/2)$ for these systems are summarized in Figure \ref{fig:Real_Systems}. The brightest source is J1720--0533 (hereafter J1720), but the distance to J1720 may be underestimated by up to a factor of 5 \citep{Koljonen_2024}. COSI has a planned line sensitivity of $1.2 \times 10^{-5} \rm ~ph/cm^2/s$ \citep{tomsick_2023}. Coherent searches for a periodic signals gain a sensitivity improvement of $\sqrt{T_{\rm obs}/P_{\rm orb}} \sim 30-300$ with $P_{\rm orb} = 10^{3}-10^{4} \text{ s}$ and $T_{\rm obs} = 10^7-10^8\text{ s}$. This improvement can be achieved without needing to time the MSP's pulsed emission directly, although timing the MSP would reduce the parameter space for the search for secondary emission. Nevertheless, J1720 would not be individually detectable with COSI, and observations would likely rely on stacking analyses of many systems similar to J1720. Given the beaming geometry of MSPs \citep{1990ApJ...352..247R,2018ApJ...863..199G}, the true source density of systems similar to J1720 (whose pulsars are not beamed toward Earth) ought to be a factor ${\cal O}(10)$ or more higher than what has been observed to date. 

In contrast with the previous section, the MSP-He WD population is considerably less bright in 511 keV line emission than the BW or RB populations. The reduction is due to the observed MSP-He WD binary systems being considerably further apart than the initial model, reducing $\Delta \Omega$, which scales as $a_{10}^{-2}$ for small $\Delta\Omega$. The observational bias against UCBs in pulsar surveys is even more so for systems with heavier companions that exacerbate Doppler smearing \citep{Pol_2021}. Rescaling our He WD model such that $P_{\rm orb} \sim 25$ min\footnote{This new geometry no longer follows the far-field source approximation used in the simulations, so the modulation with the orbit will change; this departure likely makes the "eye" shape from \textcolor{black}{Figures \ref{fig:eye} and \ref{fig:sphere}} smaller and even brighter, potentially benefiting detectability.}, which is just before the system becomes an UCXB according to \citet{Tauris_2018}, yields a peak $\dot{n}_{511}$ of $5\times10^{-9} \rm ~ph/cm^2/s$ at 100 pc and gains a coherent search sensitivity boost $\sim 80$ for COSI\textcolor{black}{, which may still be undetectable. However, an instrument like the GammaTPC concept \citep{GammaTPC} may be able to detect such a system with its estimated line sensitivity of $\sim3\times10^{-8}\mathrm{~ph/cm^2/s}$ and factoring in the sensitivity boost for a time-dependent coherent search. An even larger sensitivity boost by an additional factor of $\sim \sqrt{T_{\rm obs}/P_{\rm orb}}$  may be afforded with energy-dependent coherent searches exploiting the larger Doppler modulation of UCBs, provided a sufficient amount of photons are collected and an instrument has $\lesssim 1 \mathrm{~keV~}$ energy resolution at 511 keV. For example, a system with a time-averaged $\dot{n}_{511}$ of $10^{-9} \mathrm{~ph/cm^2/s}$ observed with an instrument with a 100 $\mathrm{cm^2}$ effective area for $10^7~\mathrm{s}$ would expect to collect only a single 511 keV photon, but a larger instrument with $\sim 10^4 \mathrm{~cm^2}$ effective area would expect to collect 100 photons per year. Yet neither COSI nor GammaTPC are expected to have the requisite energy resolution for the energy-dependent coherent searches. This may motivate development of focusing optics designed to function at 511 keV.  Detailed search prospects for COSI, GammaTPC and other instruments require dedicated simulations beyond the scope of this work, and the bias against UCBs in pulsar surveys necessitates alternative searches for these systems, not subject to the above constraints, such as with gravitational waves (GWs).} 

\subsection{MSP Binaries as LISA Sources} \label{sec:LISA}

Compact ($P_{\rm orb} \lesssim 2 \text{ days}$) MSP binary systems are GW emitters with frequencies, $f_{\rm GW} > 10^{-5}$ Hz. Following the procedure from \citet{Robson_2019} for binary systems, one can calculate the location, inclination, and polarization averaged strain spectral density for a nearly monochromatic binary system, 
\begin{equation}
    \label{eqn:h_GB}
    h_{\rm GB}(f_{\rm GW}) = \sqrt{\frac{2T_{\rm obs}\dot{f}_{\rm GW}}{3}}\frac{(GM_c/c^3)^{5/6}}{\pi^{2/3}f_{\rm GW}^{7/6}(D/c)}
\end{equation}
where $T_{\rm obs}$ is the mission lifetime, $M_{\rm c} = \left(M_{\rm NS}M_{\rm comp}\right)^{3/5}\left(M_{\rm NS}+M_{\rm comp}\right)^{-1/5}$ is the chirp mass, $D$ is the distance of the binary, and $f_{\rm GW} = 2/P_{\rm orb}$ is the GW frequency. $\dot{f}_{\rm GW}$ is given in Eq.~\ref{eqn:fdot} below for the case when GWs dominate the decay of the orbit, but non-gravitation effects such as companion mass loss and magnetic torques may alter the evolution.

\begin{equation}
    \label{eqn:fdot}
    \dot{f}_{\rm GW} = \frac{96}{5} \pi^{8/3} (G M_c/c^3)^{5/3} f_{\rm GW}^{11/3}
\end{equation}

\textcolor{black}{Equations \ref{eqn:h_GB} and \ref{eqn:fdot} together yield Equation \ref{eqn:h_GB_simplified}, where $T_{1,\mathrm{obs}} = T_\mathrm{obs}/\mathrm{yr}$, $M_{1,c} = M_c/M_\odot,$ $P_{1,\mathrm{orb}} = P_\mathrm{orb}/\mathrm{hr}$, and $D_1 = D/100 \mathrm{~pc}$}.

\begin{equation}
    \label{eqn:h_GB_simplified}
    \dot{f}_{\rm GW} = 4\times10^{-17} ~T_{1,\mathrm{obs}}^{1/2}~M_{1,c}^{5/3}~P_{1,\mathrm{orb}}^{-2/3}~D_1^{-1}
\end{equation}

As an example, we consider J1720, a BW system with a $0.034 M_\odot$ companion, $3.2$ hr orbit, and is the brightest system from Figure \ref{fig:Real_Systems}.  Assuming GW-driven orbit decay, J1720 has a $h_{\rm GB} = 9.7\times10^{-19} \, \text{Hz}^{-1/2}$, which falls near the LISA design sensitivity at $f_{\rm GW} = 0.18$ mHz for a 4 year mission. It is possible that the companion mass is underestimated for J1720 as the orbital inclination angle is poorly constrained, which would lead to an increase in GW strain amplitude.

For the models in Table \ref{tab:fluxes}, we focus on a compact binary with a He WD companion. Assuming GW-driven orbit decay, such a system would have $f_{\rm GW} = 0.17 \rm \, mHz$ and $h_{\rm GB} = 8.4\times10^{-18} \, \text{Hz}^{-1/2},$ which yields an SNR $\sim 1$ at 100 pc. Given the strong bias against pulsar UCBs, only a few similar compact binaries have been observed in a pulsar state, but LISA is sensitive to nearly the whole galaxy for those in UCXB states \citep{Tauris_2018}, and LISA searches do not experience the same degradation due to Doppler smearing \textcolor{black}{as radio pulsar surveys}. Rescaling our He WD model to just before the system becomes an UCXB according to \citet{Tauris_2018} yields a LISA SNR $\sim 200$ at 100 pc.  Conveniently, the brightest 511 keV sources are also the loudest LISA sources, so such a system could be observable with a future MeV observatory \citep[e.g. GammaTPC][]{GammaTPC}. Discovery of UCBs with LISA would define an ensemble of high-value systems for analyses with COSI and other observatories that would be unavailable with radio pulsar surveys due to Doppler smearing.

\subsection{Observable Characteristics}\label{sec:observables}

Observing the electron/positron annihilation line emission from MSP binary systems would provide a new window into the evolution of MSP systems, including constraining the NS mass and determining the composition of MSP binary companions. Suppose a nearby MSP's pulsed gamma ray emission is observed; one can measure the gamma ray luminosity from the MSP $L_{\rm MSP}$, the spectrum $\frac{dN}{dE}$, and a binary mass function (Equation \ref{eqn:Mass Function}) from the orbital period and the peak radial velocity of the MSP, $K_{\rm MSP}$. Equivalently, measuring the binary's GW strain with LISA can constrain $M_{\rm c}$, $P_{\rm orb}$, and the inclination angle. Methods for accurately identifying the MSP binary from other galactic binaries with LISA data are beyond the scope of this work.

\begin{equation}
    \label{eqn:Mass Function}
    \frac{M_{\rm comp}^3 \sin^3(i)}{M_{\rm tot}^2} = \frac{P_{\rm orb} K_{\rm MSP}^3}{2\pi G}
\end{equation}

More constraints can be placed with the additional measurements of the annihilation line flux, the flux profile over the duration of the orbit, and the Doppler shifts of the line throughout the orbit. The Doppler shifts constrain the radial velocity of the companion as Equation \ref{eqn:doppler}, which determines a second binary mass function in a similar manner to Equation \ref{eqn:Mass Function}. 

\begin{equation}
    \label{eqn:doppler}
    \frac{\Delta E}{E} = \frac{c}{c-K_{\rm comp}}
\end{equation}

Without GW measurements, the 511 keV flux variation throughout the orbit constrains the inclination angle, where face-on inclination has a constant flux, and edge-on inclination has a prominent eclipse \textcolor{black}{when the companion is between the MSP and the observer}. This variation is illustrated in \textcolor{black}{Figures \ref{fig:eye} and \ref{fig:sphere}}. From the inclination and the two mass functions or one mass function and the GW measurements, one can derive the NS and companion masses. With the masses, the semi-major axis, $a_{10}$, can be found with Equation \ref{eqn:semi-major axis}.

\begin{equation}
    \label{eqn:semi-major axis}
    G (M_{NS} + M_{c\rm omp}) = \left (\frac{2\pi}{P_{\rm orb}}\right)^2 a_{10}^3
\end{equation}

If the inclination angle is shallow enough, the line flux variation throughout the orbit also constrains $\epsilon$ by the depth of the gap between the double peak at the ecliptic. A shallower gap is correlated to more efficient production of escaped annihilation photons and to a higher-Z companion, as demonstrated with Figure \ref{fig:efficiency_dependence}. Further, if the inclination angle is small enough, then the form of $f(\phi,i = \frac{\pi}{2})$ can be approximated to find $\dot{N}_{511}$. The solid angle of the companion can be found with Equation \ref{eqn:solid angle}, where $\dot{N}_{\rm MSP}$ is the total rate of photons emitted by the MSP.

\begin{equation}
    \label{eqn:solid angle}
    \Delta \Omega = 4\pi \frac{\dot{N}_{511}}{\epsilon \dot{N}_{\rm MSP}} 
\end{equation}

Finally, the solid angle, $\Delta \Omega$, taken with the semi-major axis, a, determine the companion radius, R, as in Equation \ref{eqn:comp radius}

\begin{equation}
    \label{eqn:comp radius}
    \Delta \Omega = 2 \pi \left [ 1 - \cos(\psi)\right ] \quad , \, \,
    \tan(\psi) = \frac{R}{a}
\end{equation}

In this way, measurement of the flux of the 511 keV line combined with either pulsar timing or the binary's GW strain can constrain the composition and density of the companion in a MSP binary system.

\section{Conclusion} \label{sec:conclusion}

We introduce a new class of astrophysical 511 keV sources, (ultra-)compact binary MSP systems. We find compact He WD and massive MS binaries are the brightest companion types with $\dot{N}_{511}$ up to $10^{33}$ photons/s when the companion is the size of its Roche lobe. Given strong observational biases against pulsar UCBs, there likely exists a vast unobserved population of UCB systems with even brighter 511 keV lines. Such systems may be observable by both LISA and future soft $\gamma\text{-ray}$ instruments, providing a method to identify active pulsars whose pulsed emission is not beamed toward Earth. Among the currently observed binaries, BWs are the brightest systems. In particular, the potentially high-Z BW system, J1720, may have annihilation line emission peaking $\sim 10^{-9} \rm \, ph/cm^2/s$. Observations of the 511 keV line and its modulation constrain the orbital parameters, companion composition, and evolutionary history of the binary, opening the possibility for $\gamma\text{-ray}$ exoplanet characterization. Based on the number of MSP binaries recorded in the ATNF Pulsar Catalog \citep{Manchester_2005} within 1000 pc, we expect $0.24 \pm 0.033 \text{~(Poisson Fluctuations)~} \pm 0.077 \text{~(Beaming Fraction)}$ similar systems within 100 pc, including those whose pulsed $\gamma$-ray emission is not beamed towards Earth but not accounting for observational biases against UCBs for the observed population within 1 kpc.

A large number ($\gtrsim 10^5$) of MSPs in UCBs \citep[as proposed by some studies, e.g.,][]{2013A&A...552A..69V,10.1093/mnras/sty2135, Gautam_2022} would be required to significantly contribute to the 511 keV excess in our scenario. MSPs binaries will not account for 100\% of the 511 keV excess, because other sources such as supernova remnants and nuclear decay will contribute as well. Still even a 1\% contribution can be scientifically interesting, and a detailed population synthesis study is needed to estimate a spatially-dependent contribution by MSPs binaries. \textcolor{black}{Such a population synthesis would need to take into account both the MSP birth history in the Milky Way and their evolution as they move in the galactic potential, to obtain a credible present-day distribution of compact MSP binaries. Given the Galactic bulge has a high stellar encounter rate, additional N-body effects may be pivotal for formation of binaries. Such a study would also need to calibrate the GeV $\gamma$-ray luminosity self-consistently to {\it Fermi}-LAT observations of the Galactic center and bulge, which requires $\gtrsim 10^4$ MSPs beamed in our direction to explain the {\it Fermi}-LAT Galactic center GeV excess \citep[e.g.,][]{2018ApJ...863..199G,Gautam_2022}. In addition to a population synthesis study, other sources of primary radiation for these systems should be taken into account, as these could contribute another factor of $\gtrsim 3$ to $\dot{N}_{511}$ for some systems. Nevertheless, a rough appraisal for the contribution to the 511 keV excess can be made assuming $\dot{N}_{511}\sim10^{33}\mathrm{~ph/s}$, which of course will vary by the distribution of companion compositions, sizes, orbit compactness, and pulsar luminosities. Assuming $10^5-10^6$ such UCB MSPs (implied by \textit{Fermi}-LAT, accounting for the unseen population) in the galaxy are all located at the galactic center, their combined 511 keV flux would be $\sim 10^{-8}$ -- $10^{-7}~\mathrm{ph/cm^2/s}$. The narrow bulge flux from \citet{Skinner_20156} is about $10^{-4}~\mathrm{ph/cm^2/s}$. Thus the crude estimate here suggests a $\sim0.1\%$ is plausible for a MSP contribution to the 511 keV Galactic bulge component. However, projection effects are unaccounted for in this estimate, and a much smaller population of foreground MSPs at lower distances could significantly impact the  predicted cumulative 511 keV flux, reaching potentially a few percent or more. In \citet{Skinner_20156}, it is clear there are multiple spatially distinct components (plausibly of differing origins) that contribute to the total 511 keV flux. In this regard, COSI with its imaging capability will greatly clarify the situation.}

\textcolor{black}{Observations of nearby systems with COSI} would likely rely on joint observations with either radio pulsar surveys or LISA to define targets for stacking analyses, motivating a study of dedicated $\gamma\text{-ray}$ observatory concepts to search for this emission. Some remaining open questions are the prominence of hadronic interactions and nuclear lines due to ion bombardment (which planned instruments may be even more sensitive to than the 511 keV line), the contribution of MSP binaries to the unresolved galactic diffuse emission, the enhanced 511 keV line emission due to bombardment by relativistic species and shock emission in some binaries. Finally, such detailed studies open the possibility in the future to probe the unknown content of relativistic particle species accelerated by pulsars, and assess their relevance to cosmic rays.

\begin{acknowledgments}
We acknowledge helpful discussions with Aimee Hungerford, Peter Shawhan, Regina Caputo, and Carolyn Kierans.
The material is based upon work supported by NASA under award number 80GSFC24M0006.  This work has made use of the NASA Astrophysics Data System. 
\end{acknowledgments}

\begin{contribution}

ZM was responsible for the formal analysis, writing and submitting the manuscript. ZW came up with the initial research concept and edited the manuscript.


\end{contribution}

%

\software{MEGAlib \citep{MEGAlib}, GEANT4 \citep{GEANT4}, MESA \citep{Paxton2011,Paxton2013,Paxton2015,Paxton2018,Paxton2019,Jermyn2023}, }

\begin{table*}
    \centering
    \caption{Acronyms}
    \begin{tabular}{ll}
        ATNF        & Australia Telescope National Facility \\
        BD          & Brown Dwarf \\
        BW          & Black Widow \\
        CO          & Carbon-Oxygen \\
        COSI        & Compton Spectrometer and Imager \\
        CSDA        & Continuous Slowing Down Approximation \\
        GW          & Gravitational Wave \\
        INTEGRAL    & International Gamma-Ray Astrophysics Laboratory \\
        LAT         & Large Area Telescope \\
        LISA        & Laser Interferometer Space Antenna \\
        MS          & Main Sequence Star \\
        MSP         & Millisecond Pulsar \\
        NS          & Neutron Star \\
        RB          & Redback \\
        UCB         & Ultra-Compact Binary \\
        UCXB        & Ultra-Compact X-ray Binary \\
        UL          & Ultralight \\
        WD          & White Dwarf \\
        ZAMS        & Zero Age Main Sequence \\
        3PC         & Third Pulsar Catalog \\
    \end{tabular}
    \label{tab:placeholder}
\end{table*}


\appendix

Tables \ref{tab:real ULs1} -- \ref{tab:real Hes} list the MSP binary systems used to create Figure \ref{fig:Real_Systems}. The luminosities, companion masses, orbital periods and separations, and distances were collected from the ATNF Pulsar Catalog \citep{Manchester_2005}.

\begin{table*}
\footnotesize
\begin{ruledtabular}
\begin{tabular}{cccccccccc}
\hline
Name & 3PC & $L_{\rm MSP}$ & Mass & $\rm P_{orb}$ & Separation & Distance & Radius & Average Density & Peak Flux \\\hline
Units & -- &{ $\rm ergs/s$} & $M_\odot$ & s & cm & pc & cm & { $\rm g/cm^3$} & { $\rm ph/cm^2/s$} \\\hline
\multirow{3}{*}{J1737-0811} & \multirow{3}{*}{No} & \multirow{3}{*}{$8.6\times10^{32}$} & \multirow{3}{*}{0.076} & \multirow{3}{*}{$6.9\times10^{6}$} & \multirow{3}{*}{$6.6\times10^{12}$} & \multirow{3}{*}{206} & $2\times10^{10}$ & 10 & $1.3\times10^{-13}$ \\\cline{8-10}
 &  &  &  &  &  &  & $1\times10^{10}$ & 30 & $6.3\times10^{-14}$ \\\cline{8-10}
 &  &  &  &  &  &  & $6\times10^{8}$ & $1.7\times10^{5}$ & $2.0\times10^{-16}$ \\\hline
\multirow{3}{*}{J1502-6752} & \multirow{3}{*}{No} & \multirow{3}{*}{$1.3\times10^{32}$} & \multirow{3}{*}{0.025} & \multirow{3}{*}{$2.1\times10^{5}$} & \multirow{3}{*}{$6.5\times10^{11}$} & \multirow{3}{*}{7866} & $1\times10^{10}$ & 10 & $6.7\times10^{-16}$ \\\cline{8-10}
 &  &  &  &  &  &  & $7\times10^{9}$ & 30 & $3.2\times10^{-16}$ \\\cline{8-10}
 &  &  &  &  &  &  & $6\times10^{8}$ & $5.6\times10^{4}$ & $2.1\times10^{-18}$ \\\hline
\multirow{3}{*}{J1300+1240} & \multirow{3}{*}{No} & \multirow{3}{*}{$3.8\times10^{33}$} & \multirow{3}{*}{5.6e-08} & \multirow{3}{*}{$2.2\times10^{6}$} & \multirow{3}{*}{$3.0\times10^{12}$} & \multirow{3}{*}{709} & $1\times10^{8}$ & 10 & $1.9\times10^{-17}$ \\\cline{8-10}
 &  &  &  &  &  &  & $1\times10^{8}$ & 30 & $9.0\times10^{-18}$ \\\cline{8-10}
 &  &  &  &  &  &  & $6\times10^{8}$ & 0 & $3.5\times10^{-16}$ \\\hline
\end{tabular}
\end{ruledtabular}
\caption{\label{tab:real ULs1} The line flux estimates for the UL systems listed in the ATNF Pulsar Catalogue. We list three different assumptions for the radius and density of the companions 1) $\rm 10 ~g/cm^3$, 2) $\rm 30 ~g/cm^3$, 3) Earth radius. The values for companion mass and distance are taken from the ATNF Pulsar Catalogue \citep{Manchester_2005}. The companion mass used here is the median mass according to the ATNF Pulsar Catalogue. The separation is calculated from the orbit period listed in the ATNF Pulsar Catalogue. }
\end{table*}
\begin{table*}
\footnotesize
\begin{ruledtabular}
\begin{tabular}{cccccccccc}
\hline
Name & 3PC & $L_{\rm MSP}$ & Mass & $\rm P_{orb}$ & Separation & Distance & Radius & Average Density & Peak Flux \\\hline
Units & -- &{ $\rm ergs/s$} & $M_\odot$ & s & cm & pc & cm & { $\rm g/cm^3$} & { $\rm ph/cm^2/s$} \\\hline
J1757-5322 & No & $3.0\times10^{32}$ & 0.67 & $3.9\times10^{4}$ & $2.3\times10^{11}$ & 945 & $7\times10^{8}$ & $9.3\times10^{5}$ & $2.7\times10^{-16}$ \\\hline
J1802-2124 & No & $2.8\times10^{32}$ & 0.98 & $6.0\times10^{4}$ & $3.2\times10^{11}$ & 800 & $7\times10^{8}$ & $1.4\times10^{6}$ & $1.9\times10^{-16}$ \\\hline
J2222-0137 & No & $1.3\times10^{32}$ & 1.4 & $2.1\times10^{5}$ & $7.7\times10^{11}$ & 268 & $7\times10^{8}$ & $1.9\times10^{6}$ & $1.3\times10^{-16}$ \\\hline
J1614-2230 & Yes & $2.4\times10^{33}$ & 0.47 & $7.5\times10^{5}$ & $1.6\times10^{12}$ & 700 & $7\times10^{8}$ & $6.6\times10^{5}$ & $8.3\times10^{-17}$ \\\hline
J1952+2630 & No & $3.8\times10^{33}$ & 1.1 & $3.4\times10^{4}$ & $2.2\times10^{11}$ & 10031 & $7\times10^{8}$ & $1.6\times10^{6}$ & $3.3\times10^{-17}$ \\\hline
J1658+3630 & No & $2.5\times10^{31}$ & 1.1 & $2.6\times10^{5}$ & $8.6\times10^{11}$ & 224 & $7\times10^{8}$ & $1.5\times10^{6}$ & $3.0\times10^{-17}$ \\\hline
J1525-5545 & No & $7.1\times10^{32}$ & 0.99 & $8.6\times10^{4}$ & $4.1\times10^{11}$ & 3140 & $7\times10^{8}$ & $1.4\times10^{6}$ & $1.9\times10^{-17}$ \\\hline
J1439-5501 & No & $4.8\times10^{31}$ & 1.4 & $1.8\times10^{5}$ & $7.0\times10^{11}$ & 655 & $7\times10^{8}$ & $1.9\times10^{6}$ & $9.7\times10^{-18}$ \\\hline
J1933-6211 & No & $6.9\times10^{32}$ & 0.38 & $1.1\times10^{6}$ & $2.1\times10^{12}$ & 1000 & $7\times10^{8}$ & $5.3\times10^{5}$ & $7.1\times10^{-18}$ \\\hline
J1435-6100 & No & $2.4\times10^{32}$ & 1.1 & $1.2\times10^{5}$ & $5.0\times10^{11}$ & 2816 & $7\times10^{8}$ & $1.5\times10^{6}$ & $5.1\times10^{-18}$ \\\hline
J1045-0436 & No & $4.5\times10^{31}$ & 0.98 & $8.9\times10^{5}$ & $1.9\times10^{12}$ & 329 & $7\times10^{8}$ & $1.4\times10^{6}$ & $4.8\times10^{-18}$ \\\hline
J2053+4650 & No & $6.8\times10^{32}$ & 1.0 & $2.1\times10^{5}$ & $7.4\times10^{11}$ & 3810 & $7\times10^{8}$ & $1.4\times10^{6}$ & $3.7\times10^{-18}$ \\\hline
J2145-0750 & No & $5.7\times10^{31}$ & 0.5 & $5.9\times10^{5}$ & $1.4\times10^{12}$ & 625 & $7\times10^{8}$ & $7.0\times10^{5}$ & $3.3\times10^{-18}$ \\\hline
J1022+1001 & No & $7.7\times10^{31}$ & 0.85 & $6.7\times10^{5}$ & $1.6\times10^{12}$ & 725 & $7\times10^{8}$ & $1.2\times10^{6}$ & $2.5\times10^{-18}$ \\\hline
J0900-3144 & No & $2.8\times10^{32}$ & 0.42 & $1.6\times10^{6}$ & $2.7\times10^{12}$ & 900 & $7\times10^{8}$ & $5.9\times10^{5}$ & $2.1\times10^{-18}$ \\\hline
J0824+0028 & No & $1.2\times10^{33}$ & 0.4 & $2.0\times10^{6}$ & $3.1\times10^{12}$ & 1689 & $7\times10^{8}$ & $5.6\times10^{5}$ & $2.0\times10^{-18}$ \\\hline
J1949+3106 & No & $3.3\times10^{32}$ & 0.97 & $1.7\times10^{5}$ & $6.3\times10^{11}$ & 7468 & $7\times10^{8}$ & $1.3\times10^{6}$ & $6.3\times10^{-19}$ \\\hline
J0721-2038 & No & $9.3\times10^{31}$ & 0.55 & $4.7\times10^{5}$ & $1.2\times10^{12}$ & 2680 & $7\times10^{8}$ & $7.6\times10^{5}$ & $3.9\times10^{-19}$ \\\hline
J1101-6424 & No & $1.1\times10^{32}$ & 0.57 & $8.3\times10^{5}$ & $1.7\times10^{12}$ & 2175 & $7\times10^{8}$ & $7.9\times10^{5}$ & $3.2\times10^{-19}$ \\\hline
J1943+2210 & No & $5.3\times10^{32}$ & 0.33 & $7.2\times10^{5}$ & $1.5\times10^{12}$ & 6773 & $7\times10^{8}$ & $4.6\times10^{5}$ & $2.2\times10^{-19}$ \\\hline
J1337-6423 & No & $2.3\times10^{32}$ & 0.95 & $4.1\times10^{5}$ & $1.2\times10^{12}$ & 5951 & $7\times10^{8}$ & $1.3\times10^{6}$ & $2.1\times10^{-19}$ \\\hline
J1938+6604 & No & $1.4\times10^{31}$ & 1.0 & $2.1\times10^{5}$ & $7.5\times10^{11}$ & 3379 & $7\times10^{8}$ & $1.4\times10^{6}$ & $9.4\times10^{-20}$ \\\hline
J1603-7202 & No & $3.8\times10^{31}$ & 0.34 & $5.5\times10^{5}$ & $1.3\times10^{12}$ & 3700 & $7\times10^{8}$ & $4.7\times10^{5}$ & $7.4\times10^{-20}$ \\\hline
J1933+1726 & No & $3.9\times10^{31}$ & 0.94 & $4.5\times10^{5}$ & $1.2\times10^{12}$ & 4072 & $7\times10^{8}$ & $1.3\times10^{6}$ & $6.9\times10^{-20}$ \\\hline
J1755-3716 & No & $1.2\times10^{32}$ & 0.35 & $9.9\times10^{5}$ & $1.9\times10^{12}$ & 8178 & $7\times10^{8}$ & $4.9\times10^{5}$ & $2.1\times10^{-20}$ \\\hline
\end{tabular}
\end{ruledtabular}
\caption{\label{tab:real COs} The line flux estimates for C WD-MSP systems in the ATNF Pulsar Catalogue. The values for companion mass and distance are taken from the ATNF Pulsar Catalogue \citep{Manchester_2005}. The companion mass used here is the median mass according to the ATNF Pulsar Catalogue. The separation is calculated from the orbit period listed in the ATNF Pulsar Catalogue. The radius is the lesser value of either the size of the Roche lobe or the radius of the model used in the Methods section. The density column is the average density assuming a homogeneous sphere with the companion mass and radius.}
\end{table*}
\begin{table*}
\footnotesize
\begin{ruledtabular}
\begin{tabular}{ccccccccccc}
\hline
Name & 3PC & $L_{\rm MSP}$ & Mass & $\rm P_{orb}$ & Separation & Distance & Radius & Average Density & Model & Peak Flux \\\hline
Units & -- &{ $\rm ergs/s$} & $M_\odot$ & s & cm & pc & cm & { $\rm g/cm^3$} & -- & { $\rm ph/cm^2/s$} \\\hline
J2339-0533 & Yes & $4.6\times10^{33}$ & 0.3 & $1.7\times10^{4}$ & $1.2\times10^{11}$ & 1100 & $2\times10^{10}$ & 18 & 0.3 MS & $1.2\times10^{-12}$ \\\hline
J2215+5135 & Yes & $1.5\times10^{34}$ & 0.24 & $1.5\times10^{4}$ & $1.1\times10^{11}$ & 2773 & $2\times10^{10}$ & 15 & 0.3 MS & $6.8\times10^{-13}$ \\\hline
J1431-4715 & Yes & $1.4\times10^{34}$ & 0.14 & $3.9\times10^{4}$ & $2.1\times10^{11}$ & 1562 & $8\times10^{9}$ & $1.2\times10^{2}$ & 0.1 MS & $9.7\times10^{-14}$ \\\hline
J1816+4510 & Yes & $1.0\times10^{34}$ & 0.18 & $3.1\times10^{4}$ & $1.8\times10^{11}$ & 4356 & $8\times10^{9}$ & $1.5\times10^{2}$ & 0.1 MS & $1.3\times10^{-14}$ \\\hline
\end{tabular}
\end{ruledtabular}
\caption{\label{tab:real MSs} The line flux estimates for MS star-MSP systems in the ATNF Pulsar Catalogue. The values for companion mass and distance are taken from the ATNF Pulsar Catalogue \citep{Manchester_2005}. The companion mass used here is the median mass according to the ATNF Pulsar Catalogue. The separation is calculated from the orbit period listed in the ATNF Pulsar Catalogue. The radius is the lesser value of either the size of the Roche lobe or the radius of the model used in the Methods section. The density column is the average density assuming a homogeneous sphere with the companion mass and radius. The model column lists which model from the Methods section is used.}
\end{table*}
\clearpage
\begin{longtable*}{cccccccccc}
\footnotesize
Name & 3PC & $L_{\rm MSP}$ & Mass & $\rm P_{orb}$ & Separation & Distance & Radius & Average Density & Peak Flux \\\hline
Units & -- &{ $\rm ergs/s$} & $M_\odot$ & s & cm & pc & cm & { $\rm g/cm^3$} & { $\rm ph/cm^2/s$} \\\hline

\endhead
\hline
\hline
\hline
\hline
\endfoot
\endlastfoot
\hline
\multirow{3}{*}{J1720-0533} & \multirow{3}{*}{No} & \multirow{3}{*}{$1.8\times10^{33}$} & \multirow{3}{*}{0.034} & \multirow{3}{*}{$1.1\times10^{4}$} & \multirow{3}{*}{$9.1\times10^{10}$} & \multirow{3}{*}{191} & $1\times10^{10}$ & 10 & $9.4\times10^{-10}$ \\\cline{8-10}
 &  &  &  &  &  &  & $8\times10^{9}$ & 30 & $4.7\times10^{-10}$ \\\cline{8-10}
 &  &  &  &  &  &  & $6\times10^{8}$ & $7.5\times10^{4}$ & $2.6\times10^{-12}$ \\\hline
\multirow{3}{*}{J1959+2048} & \multirow{3}{*}{Yes} & \multirow{3}{*}{$3.2\times10^{34}$} & \multirow{3}{*}{0.025} & \multirow{3}{*}{$3.3\times10^{4}$} & \multirow{3}{*}{$1.9\times10^{11}$} & \multirow{3}{*}{1400} & $1\times10^{10}$ & 10 & $6.3\times10^{-11}$ \\\cline{8-10}
 &  &  &  &  &  &  & $7\times10^{9}$ & 30 & $3.0\times10^{-11}$ \\\cline{8-10}
 &  &  &  &  &  &  & $6\times10^{8}$ & $5.5\times10^{4}$ & $2.0\times10^{-13}$ \\\hline
\multirow{3}{*}{J2241-5236} & \multirow{3}{*}{Yes} & \multirow{3}{*}{$5.2\times10^{33}$} & \multirow{3}{*}{0.014} & \multirow{3}{*}{$1.3\times10^{4}$} & \multirow{3}{*}{$9.7\times10^{10}$} & \multirow{3}{*}{1042} & $9\times10^{9}$ & 10 & $4.5\times10^{-11}$ \\\cline{8-10}
 &  &  &  &  &  &  & $6\times10^{9}$ & 30 & $2.1\times10^{-11}$ \\\cline{8-10}
 &  &  &  &  &  &  & $6\times10^{8}$ & $3.0\times10^{4}$ & $2.2\times10^{-13}$ \\\hline
\multirow{3}{*}{J1701-3006F} & \multirow{3}{*}{No} & \multirow{3}{*}{$1.5\times10^{35}$} & \multirow{3}{*}{0.024} & \multirow{3}{*}{$1.8\times10^{4}$} & \multirow{3}{*}{$1.2\times10^{11}$} & \multirow{3}{*}{6410} & $1\times10^{10}$ & 10 & $3.0\times10^{-11}$ \\\cline{8-10}
 &  &  &  &  &  &  & $7\times10^{9}$ & 30 & $1.5\times10^{-11}$ \\\cline{8-10}
 &  &  &  &  &  &  & $6\times10^{8}$ & $5.3\times10^{4}$ & $1.0\times10^{-13}$ \\\hline
\multirow{3}{*}{J1701-3006E} & \multirow{3}{*}{No} & \multirow{3}{*}{$7.2\times10^{34}$} & \multirow{3}{*}{0.035} & \multirow{3}{*}{$1.4\times10^{4}$} & \multirow{3}{*}{$1.0\times10^{11}$} & \multirow{3}{*}{6410} & $1\times10^{10}$ & 10 & $2.7\times10^{-11}$ \\\cline{8-10}
 &  &  &  &  &  &  & $8\times10^{9}$ & 30 & $1.3\times10^{-11}$ \\\cline{8-10}
 &  &  &  &  &  &  & $6\times10^{8}$ & $7.8\times10^{4}$ & $7.0\times10^{-14}$ \\\hline
\multirow{3}{*}{J2214+3000} & \multirow{3}{*}{Yes} & \multirow{3}{*}{$3.8\times10^{33}$} & \multirow{3}{*}{0.015} & \multirow{3}{*}{$3.6\times10^{4}$} & \multirow{3}{*}{$2.0\times10^{11}$} & \multirow{3}{*}{600} & $9\times10^{9}$ & 10 & $2.7\times10^{-11}$ \\\cline{8-10}
 &  &  &  &  &  &  & $6\times10^{9}$ & 30 & $1.3\times10^{-11}$ \\\cline{8-10}
 &  &  &  &  &  &  & $6\times10^{8}$ & $3.4\times10^{4}$ & $1.2\times10^{-13}$ \\\hline
\multirow{3}{*}{J0251+2606} & \multirow{3}{*}{Yes} & \multirow{3}{*}{$3.6\times10^{33}$} & \multirow{3}{*}{0.028} & \multirow{3}{*}{$1.7\times10^{4}$} & \multirow{3}{*}{$1.2\times10^{11}$} & \multirow{3}{*}{1170} & $1\times10^{10}$ & 10 & $2.6\times10^{-11}$ \\\cline{8-10}
 &  &  &  &  &  &  & $8\times10^{9}$ & 30 & $1.2\times10^{-11}$ \\\cline{8-10}
 &  &  &  &  &  &  & $6\times10^{8}$ & $6.2\times10^{4}$ & $7.7\times10^{-14}$ \\\hline
\multirow{3}{*}{J0024-7204R} & \multirow{3}{*}{No} & \multirow{3}{*}{$2.8\times10^{34}$} & \multirow{3}{*}{0.03} & \multirow{3}{*}{$5.7\times10^{3}$} & \multirow{3}{*}{$5.8\times10^{10}$} & \multirow{3}{*}{4520} & $7\times10^{9}$ & 41 & $2.4\times10^{-11}$ \\\cline{8-10}
 &  &  &  &  &  &  & $7\times10^{9}$ & 41 & $2.4\times10^{-11}$ \\\cline{8-10}
 &  &  &  &  &  &  & $6\times10^{8}$ & $6.6\times10^{4}$ & $1.7\times10^{-13}$ \\\hline
\multirow{3}{*}{J1641+8049} & \multirow{3}{*}{Yes} & \multirow{3}{*}{$8.6\times10^{33}$} & \multirow{3}{*}{0.047} & \multirow{3}{*}{$7.9\times10^{3}$} & \multirow{3}{*}{$7.1\times10^{10}$} & \multirow{3}{*}{3035} & $1\times10^{10}$ & 22 & $2.1\times10^{-11}$ \\\cline{8-10}
 &  &  &  &  &  &  & $9\times10^{9}$ & 30 & $1.7\times10^{-11}$ \\\cline{8-10}
 &  &  &  &  &  &  & $6\times10^{8}$ & $1.0\times10^{5}$ & $7.7\times10^{-14}$ \\\hline
\multirow{3}{*}{J2234+0944} & \multirow{3}{*}{Yes} & \multirow{3}{*}{$3.3\times10^{33}$} & \multirow{3}{*}{0.018} & \multirow{3}{*}{$3.6\times10^{4}$} & \multirow{3}{*}{$2.0\times10^{11}$} & \multirow{3}{*}{714} & $9\times10^{9}$ & 10 & $1.8\times10^{-11}$ \\\cline{8-10}
 &  &  &  &  &  &  & $7\times10^{9}$ & 30 & $8.5\times10^{-12}$ \\\cline{8-10}
 &  &  &  &  &  &  & $6\times10^{8}$ & $3.9\times10^{4}$ & $7.1\times10^{-14}$ \\\hline
\multirow{3}{*}{J0636+5128} & \multirow{3}{*}{Yes} & \multirow{3}{*}{$1.2\times10^{33}$} & \multirow{3}{*}{0.0079} & \multirow{3}{*}{$5.8\times10^{3}$} & \multirow{3}{*}{$5.8\times10^{10}$} & \multirow{3}{*}{714} & $5\times10^{9}$ & 39 & $1.7\times10^{-11}$ \\\cline{8-10}
 &  &  &  &  &  &  & $5\times10^{9}$ & 39 & $1.7\times10^{-11}$ \\\cline{8-10}
 &  &  &  &  &  &  & $6\times10^{8}$ & $1.8\times10^{4}$ & $2.9\times10^{-13}$ \\\hline
\multirow{3}{*}{J1446-4701} & \multirow{3}{*}{Yes} & \multirow{3}{*}{$7.3\times10^{33}$} & \multirow{3}{*}{0.022} & \multirow{3}{*}{$2.4\times10^{4}$} & \multirow{3}{*}{$1.5\times10^{11}$} & \multirow{3}{*}{1569} & $1\times10^{10}$ & 10 & $1.6\times10^{-11}$ \\\cline{8-10}
 &  &  &  &  &  &  & $7\times10^{9}$ & 30 & $7.8\times10^{-12}$ \\\cline{8-10}
 &  &  &  &  &  &  & $6\times10^{8}$ & $4.9\times10^{4}$ & $5.6\times10^{-14}$ \\\hline
\multirow{3}{*}{J1555-2908} & \multirow{3}{*}{Yes} & \multirow{3}{*}{$6.2\times10^{34}$} & \multirow{3}{*}{0.059} & \multirow{3}{*}{$2.0\times10^{4}$} & \multirow{3}{*}{$1.3\times10^{11}$} & \multirow{3}{*}{7559} & $1\times10^{10}$ & 10 & $1.4\times10^{-11}$ \\\cline{8-10}
 &  &  &  &  &  &  & $1\times10^{10}$ & 30 & $6.8\times10^{-12}$ \\\cline{8-10}
 &  &  &  &  &  &  & $6\times10^{8}$ & $1.3\times10^{5}$ & $2.5\times10^{-14}$ \\\hline
\multirow{3}{*}{J0023+0923} & \multirow{3}{*}{Yes} & \multirow{3}{*}{$3.2\times10^{33}$} & \multirow{3}{*}{0.019} & \multirow{3}{*}{$1.2\times10^{4}$} & \multirow{3}{*}{$9.4\times10^{10}$} & \multirow{3}{*}{1818} & $1\times10^{10}$ & 10 & $1.2\times10^{-11}$ \\\cline{8-10}
 &  &  &  &  &  &  & $7\times10^{9}$ & 30 & $5.7\times10^{-12}$ \\\cline{8-10}
 &  &  &  &  &  &  & $6\times10^{8}$ & $4.2\times10^{4}$ & $4.6\times10^{-14}$ \\\hline
\multirow{3}{*}{J1221-0633} & \multirow{3}{*}{Yes} & \multirow{3}{*}{$5.7\times10^{33}$} & \multirow{3}{*}{0.015} & \multirow{3}{*}{$3.3\times10^{4}$} & \multirow{3}{*}{$1.9\times10^{11}$} & \multirow{3}{*}{1251} & $9\times10^{9}$ & 10 & $1.0\times10^{-11}$ \\\cline{8-10}
 &  &  &  &  &  &  & $6\times10^{9}$ & 30 & $4.8\times10^{-12}$ \\\cline{8-10}
 &  &  &  &  &  &  & $6\times10^{8}$ & $3.4\times10^{4}$ & $4.5\times10^{-14}$ \\\hline
\multirow{3}{*}{J1957+2516} & \multirow{3}{*}{No} & \multirow{3}{*}{$3.5\times10^{33}$} & \multirow{3}{*}{0.11} & \multirow{3}{*}{$2.1\times10^{4}$} & \multirow{3}{*}{$1.4\times10^{11}$} & \multirow{3}{*}{2659} & $2\times10^{10}$ & 10 & $9.3\times10^{-12}$ \\\cline{8-10}
 &  &  &  &  &  &  & $1\times10^{10}$ & 30 & $4.5\times10^{-12}$ \\\cline{8-10}
 &  &  &  &  &  &  & $6\times10^{8}$ & $2.5\times10^{5}$ & $1.1\times10^{-14}$ \\\hline
\multirow{3}{*}{J2051-0827} & \multirow{3}{*}{Yes} & \multirow{3}{*}{$1.1\times10^{33}$} & \multirow{3}{*}{0.031} & \multirow{3}{*}{$8.6\times10^{3}$} & \multirow{3}{*}{$7.5\times10^{10}$} & \multirow{3}{*}{1469} & $9\times10^{9}$ & 18 & $9.0\times10^{-12}$ \\\cline{8-10}
 &  &  &  &  &  &  & $8\times10^{9}$ & 30 & $6.5\times10^{-12}$ \\\cline{8-10}
 &  &  &  &  &  &  & $6\times10^{8}$ & $6.8\times10^{4}$ & $3.8\times10^{-14}$ \\\hline
\multirow{3}{*}{J1719-1438} & \multirow{3}{*}{No} & \multirow{3}{*}{$3.3\times10^{32}$} & \multirow{3}{*}{0.0013} & \multirow{3}{*}{$7.8\times10^{3}$} & \multirow{3}{*}{$7.1\times10^{10}$} & \multirow{3}{*}{336} & $3\times10^{9}$ & 20 & $6.7\times10^{-12}$ \\\cline{8-10}
 &  &  &  &  &  &  & $3\times10^{9}$ & 30 & $5.2\times10^{-12}$ \\\cline{8-10}
 &  &  &  &  &  &  & $6\times10^{8}$ & $2.9\times10^{3}$ & $2.5\times10^{-13}$ \\\hline
\multirow{3}{*}{J1630+3550} & \multirow{3}{*}{No} & \multirow{3}{*}{$4.9\times10^{33}$} & \multirow{3}{*}{0.011} & \multirow{3}{*}{$2.7\times10^{4}$} & \multirow{3}{*}{$1.6\times10^{11}$} & \multirow{3}{*}{1568} & $8\times10^{9}$ & 10 & $5.9\times10^{-12}$ \\\cline{8-10}
 &  &  &  &  &  &  & $6\times10^{9}$ & 30 & $2.8\times10^{-12}$ \\\cline{8-10}
 &  &  &  &  &  &  & $6\times10^{8}$ & $2.5\times10^{4}$ & $3.2\times10^{-14}$ \\\hline
\multirow{3}{*}{J1805+0615} & \multirow{3}{*}{Yes} & \multirow{3}{*}{$1.9\times10^{34}$} & \multirow{3}{*}{0.027} & \multirow{3}{*}{$2.9\times10^{4}$} & \multirow{3}{*}{$1.7\times10^{11}$} & \multirow{3}{*}{3885} & $1\times10^{10}$ & 10 & $5.8\times10^{-12}$ \\\cline{8-10}
 &  &  &  &  &  &  & $8\times10^{9}$ & 30 & $2.8\times10^{-12}$ \\\cline{8-10}
 &  &  &  &  &  &  & $6\times10^{8}$ & $5.9\times10^{4}$ & $1.8\times10^{-14}$ \\\hline
\multirow{3}{*}{J0610-2100} & \multirow{3}{*}{Yes} & \multirow{3}{*}{$1.7\times10^{33}$} & \multirow{3}{*}{0.025} & \multirow{3}{*}{$2.5\times10^{4}$} & \multirow{3}{*}{$1.5\times10^{11}$} & \multirow{3}{*}{1389} & $1\times10^{10}$ & 10 & $4.9\times10^{-12}$ \\\cline{8-10}
 &  &  &  &  &  &  & $7\times10^{9}$ & 30 & $2.4\times10^{-12}$ \\\cline{8-10}
 &  &  &  &  &  &  & $6\times10^{8}$ & $5.5\times10^{4}$ & $1.6\times10^{-14}$ \\\hline
\multirow{3}{*}{J1731-1847} & \multirow{3}{*}{No} & \multirow{3}{*}{$1.6\times10^{34}$} & \multirow{3}{*}{0.039} & \multirow{3}{*}{$2.7\times10^{4}$} & \multirow{3}{*}{$1.6\times10^{11}$} & \multirow{3}{*}{4782} & $1\times10^{10}$ & 10 & $4.6\times10^{-12}$ \\\cline{8-10}
 &  &  &  &  &  &  & $8\times10^{9}$ & 30 & $2.2\times10^{-12}$ \\\cline{8-10}
 &  &  &  &  &  &  & $6\times10^{8}$ & $8.5\times10^{4}$ & $1.1\times10^{-14}$ \\\hline
\multirow{3}{*}{J1544+4937} & \multirow{3}{*}{Yes} & \multirow{3}{*}{$2.2\times10^{33}$} & \multirow{3}{*}{0.02} & \multirow{3}{*}{$1.0\times10^{4}$} & \multirow{3}{*}{$8.6\times10^{10}$} & \multirow{3}{*}{2990} & $9\times10^{9}$ & 12 & $3.3\times10^{-12}$ \\\cline{8-10}
 &  &  &  &  &  &  & $7\times10^{9}$ & 30 & $1.8\times10^{-12}$ \\\cline{8-10}
 &  &  &  &  &  &  & $6\times10^{8}$ & $4.3\times10^{4}$ & $1.4\times10^{-14}$ \\\hline
\multirow{3}{*}{J1641+3627E} & \multirow{3}{*}{No} & \multirow{3}{*}{$9.0\times10^{33}$} & \multirow{3}{*}{0.023} & \multirow{3}{*}{$9.7\times10^{3}$} & \multirow{3}{*}{$8.2\times10^{10}$} & \multirow{3}{*}{7500} & $9\times10^{9}$ & 14 & $2.4\times10^{-12}$ \\\cline{8-10}
 &  &  &  &  &  &  & $7\times10^{9}$ & 30 & $1.4\times10^{-12}$ \\\cline{8-10}
 &  &  &  &  &  &  & $6\times10^{8}$ & $5.1\times10^{4}$ & $1.0\times10^{-14}$ \\\hline
\multirow{3}{*}{J1745+1017} & \multirow{3}{*}{Yes} & \multirow{3}{*}{$1.2\times10^{33}$} & \multirow{3}{*}{0.016} & \multirow{3}{*}{$6.3\times10^{4}$} & \multirow{3}{*}{$2.8\times10^{11}$} & \multirow{3}{*}{1214} & $9\times10^{9}$ & 10 & $9.5\times10^{-13}$ \\\cline{8-10}
 &  &  &  &  &  &  & $6\times10^{9}$ & 30 & $4.5\times10^{-13}$ \\\cline{8-10}
 &  &  &  &  &  &  & $6\times10^{8}$ & $3.5\times10^{4}$ & $4.1\times10^{-15}$ \\\hline
\multirow{3}{*}{J2055+3829} & \multirow{3}{*}{No} & \multirow{3}{*}{$8.7\times10^{32}$} & \multirow{3}{*}{0.026} & \multirow{3}{*}{$1.1\times10^{4}$} & \multirow{3}{*}{$9.0\times10^{10}$} & \multirow{3}{*}{4588} & $1\times10^{10}$ & 11 & $6.5\times10^{-13}$ \\\cline{8-10}
 &  &  &  &  &  &  & $7\times10^{9}$ & 30 & $3.3\times10^{-13}$ \\\cline{8-10}
 &  &  &  &  &  &  & $6\times10^{8}$ & $5.7\times10^{4}$ & $2.2\times10^{-15}$ \\\hline
\multirow{3}{*}{J1836-2354A} & \multirow{3}{*}{No} & \multirow{3}{*}{$4.8\times10^{32}$} & \multirow{3}{*}{0.02} & \multirow{3}{*}{$1.8\times10^{4}$} & \multirow{3}{*}{$1.2\times10^{11}$} & \multirow{3}{*}{3300} & $1\times10^{10}$ & 10 & $3.4\times10^{-13}$ \\\cline{8-10}
 &  &  &  &  &  &  & $7\times10^{9}$ & 30 & $1.6\times10^{-13}$ \\\cline{8-10}
 &  &  &  &  &  &  & $6\times10^{8}$ & $4.3\times10^{4}$ & $1.3\times10^{-15}$ \\\hline
\multirow{3}{*}{J1850+0242} & \multirow{3}{*}{No} & \multirow{3}{*}{$1.4\times10^{34}$} & \multirow{3}{*}{0.083} & \multirow{3}{*}{$6.4\times10^{4}$} & \multirow{3}{*}{$2.9\times10^{11}$} & \multirow{3}{*}{12311} & $2\times10^{10}$ & 10 & $3.3\times10^{-13}$ \\\cline{8-10}
 &  &  &  &  &  &  & $1\times10^{10}$ & 30 & $1.6\times10^{-13}$ \\\cline{8-10}
 &  &  &  &  &  &  & $6\times10^{8}$ & $1.8\times10^{5}$ & $4.8\times10^{-16}$ \\\hline
\multirow{3}{*}{J1317-0157} & \multirow{3}{*}{No} & \multirow{3}{*}{$1.8\times10^{33}$} & \multirow{3}{*}{0.02} & \multirow{3}{*}{$7.7\times10^{3}$} & \multirow{3}{*}{$7.0\times10^{10}$} & \multirow{3}{*}{25000} & $8\times10^{9}$ & 22 & $3.8\times10^{-14}$ \\\cline{8-10}
 &  &  &  &  &  &  & $7\times10^{9}$ & 30 & $3.2\times10^{-14}$ \\\cline{8-10}
 &  &  &  &  &  &  & $6\times10^{8}$ & $4.5\times10^{4}$ & $2.4\times10^{-16}$ \\\hline
\multirow{3}{*}{J2322-2650} & \multirow{3}{*}{No} & \multirow{3}{*}{$1.1\times10^{32}$} & \multirow{3}{*}{0.00085} & \multirow{3}{*}{$2.8\times10^{4}$} & \multirow{3}{*}{$1.6\times10^{11}$} & \multirow{3}{*}{2000} & $3\times10^{9}$ & 10 & $1.4\times10^{-14}$ \\\cline{8-10}
 &  &  &  &  &  &  & $2\times10^{9}$ & 30 & $6.9\times10^{-15}$ \\\cline{8-10}
 &  &  &  &  &  &  & $6\times10^{8}$ & $1.9\times10^{3}$ & $4.3\times10^{-16}$ \\\hline
\caption{\label{tab:real BWs1} The line flux estimates for the BW systems listed in the ATNF Pulsar Catalogue. We list three different assumptions for the radius and density of the companions 1) $\rm 10 ~g/cm^3$, 2) $\rm 30 ~g/cm^3$, 3) Earth radius. The values for companion mass and distance are taken from the ATNF Pulsar Catalogue \citep{Manchester_2005}. The companion mass used here is the median mass according to the ATNF Pulsar Catalogue. The separation is calculated from the orbit period listed in the ATNF Pulsar Catalogue. }
\end{longtable*}
\clearpage
\null
\clearpage
\begin{longtable*}{cccccccccc}
\footnotesize
Name & 3PC & $L_{\rm MSP}$ & Mass & $\rm P_{orb}$ & Separation & Distance & Radius & Average Density & Peak Flux \\\hline
Units & -- & { $\rm ergs/s$} & $M_\odot$ & s & cm & pc & cm & { $\rm g/cm^3$} & { $\rm ph/cm^2/s$} \\
\endhead
\hline
\hline
\hline
\hline
\endfoot
\endlastfoot
\hline
J0751+1807 & Yes & $1.5\times10^{33}$ & 0.15 & $2.3\times10^{4}$ & $1.5\times10^{11}$ & 600 & $2\times10^{10}$ & 9 & $1.5\times10^{-12}$ \\\hline
J0437-4715 & Yes & $2.4\times10^{33}$ & 0.16 & $5.0\times10^{5}$ & $1.2\times10^{12}$ & 157 & $2\times10^{10}$ & 10 & $5.8\times10^{-13}$ \\\hline
J1231-1411 & Yes & $3.6\times10^{33}$ & 0.22 & $1.6\times10^{5}$ & $5.5\times10^{11}$ & 420 & $2\times10^{10}$ & 13 & $5.3\times10^{-13}$ \\\hline
J0742+4110 & No & $1.7\times10^{33}$ & 0.067 & $1.2\times10^{5}$ & $4.4\times10^{11}$ & 523 & $2\times10^{10}$ & 4 & $2.6\times10^{-13}$ \\\hline
J0621+2514 & Yes & $9.7\times10^{33}$ & 0.17 & $1.1\times10^{5}$ & $4.2\times10^{11}$ & 1641 & $2\times10^{10}$ & 10 & $1.6\times10^{-13}$ \\\hline
J0613-0200 & Yes & $2.6\times10^{33}$ & 0.15 & $1.0\times10^{5}$ & $4.1\times10^{11}$ & 950 & $2\times10^{10}$ & 9 & $1.4\times10^{-13}$ \\\hline
J0337+1715 & No & $6.8\times10^{33}$ & 0.14 & $1.4\times10^{5}$ & $5.0\times10^{11}$ & 1300 & $2\times10^{10}$ & 8 & $1.3\times10^{-13}$ \\\hline
J1902-5105 & Yes & $1.4\times10^{34}$ & 0.19 & $1.7\times10^{5}$ & $5.8\times10^{11}$ & 1645 & $2\times10^{10}$ & 11 & $1.2\times10^{-13}$ \\\hline
J0218+4232 & Yes & $4.9\times10^{34}$ & 0.2 & $1.8\times10^{5}$ & $5.8\times10^{11}$ & 3150 & $2\times10^{10}$ & 12 & $1.2\times10^{-13}$ \\\hline
J1909-3744 & Yes & $4.3\times10^{33}$ & 0.23 & $1.3\times10^{5}$ & $4.9\times10^{11}$ & 1140 & $2\times10^{10}$ & 14 & $1.1\times10^{-13}$ \\\hline
J0034-0534 & Yes & $5.9\times10^{33}$ & 0.16 & $1.4\times10^{5}$ & $4.9\times10^{11}$ & 1348 & $2\times10^{10}$ & 10 & $1.1\times10^{-13}$ \\\hline
J1514-4946 & Yes & $3.2\times10^{33}$ & 0.2 & $1.7\times10^{5}$ & $5.6\times10^{11}$ & 908 & $2\times10^{10}$ & 12 & $9.8\times10^{-14}$ \\\hline
J0024-7204U & No & $9.2\times10^{33}$ & 0.14 & $3.7\times10^{4}$ & $2.0\times10^{11}$ & 4520 & $2\times10^{10}$ & 9 & $8.5\times10^{-14}$ \\\hline
J1738+0333 & No & $9.5\times10^{32}$ & 0.1 & $3.1\times10^{4}$ & $1.8\times10^{11}$ & 1695 & $2\times10^{10}$ & 6 & $8.2\times10^{-14}$ \\\hline
J1400-1431 & Yes & $1.9\times10^{33}$ & 0.31 & $8.2\times10^{5}$ & $1.7\times10^{12}$ & 278 & $2\times10^{10}$ & 19 & $7.3\times10^{-14}$ \\\hline
J0101-6422 & Yes & $2.4\times10^{33}$ & 0.18 & $1.5\times10^{5}$ & $5.3\times10^{11}$ & 1001 & $2\times10^{10}$ & 11 & $6.7\times10^{-14}$ \\\hline
J2043+1711 & Yes & $3.1\times10^{33}$ & 0.2 & $1.3\times10^{5}$ & $4.7\times10^{11}$ & 1389 & $2\times10^{10}$ & 12 & $5.7\times10^{-14}$ \\\hline
J2006+0148 & Yes & $2.6\times10^{33}$ & 0.18 & $5.6\times10^{4}$ & $2.7\times10^{11}$ & 2437 & $2\times10^{10}$ & 10 & $4.7\times10^{-14}$ \\\hline
J1543-5149 & Yes & $1.5\times10^{34}$ & 0.26 & $7.0\times10^{5}$ & $1.5\times10^{12}$ & 1149 & $2\times10^{10}$ & 16 & $4.1\times10^{-14}$ \\\hline
J1625-0021 & Yes & $7.4\times10^{33}$ & 0.19 & $6.4\times10^{5}$ & $1.4\times10^{12}$ & 951 & $2\times10^{10}$ & 11 & $3.5\times10^{-14}$ \\\hline
J1911-1114 & No & $2.3\times10^{33}$ & 0.14 & $2.3\times10^{5}$ & $7.0\times10^{11}$ & 1069 & $2\times10^{10}$ & 8 & $3.3\times10^{-14}$ \\\hline
J1701-3006D & No & $2.5\times10^{34}$ & 0.14 & $9.7\times10^{4}$ & $3.9\times10^{11}$ & 6410 & $2\times10^{10}$ & 8 & $3.2\times10^{-14}$ \\\hline
J1835-3259B & Yes & $5.6\times10^{34}$ & 0.2 & $1.0\times10^{5}$ & $4.1\times10^{11}$ & 9460 & $2\times10^{10}$ & 12 & $3.0\times10^{-14}$ \\\hline
J2017+0603 & Yes & $2.6\times10^{33}$ & 0.21 & $1.9\times10^{5}$ & $6.1\times10^{11}$ & 1399 & $2\times10^{10}$ & 12 & $2.8\times10^{-14}$ \\\hline
J1045-4509 & No & $3.3\times10^{32}$ & 0.19 & $3.5\times10^{5}$ & $9.3\times10^{11}$ & 340 & $2\times10^{10}$ & 11 & $2.7\times10^{-14}$ \\\hline
J0740+6620 & Yes & $4.0\times10^{33}$ & 0.22 & $4.1\times10^{5}$ & $1.0\times10^{12}$ & 1150 & $2\times10^{10}$ & 13 & $2.3\times10^{-14}$ \\\hline
J1745-0952 & No & $1.0\times10^{32}$ & 0.13 & $4.3\times10^{5}$ & $1.0\times10^{12}$ & 226 & $2\times10^{10}$ & 7 & $1.5\times10^{-14}$ \\\hline
J0024-7204T & No & $5.3\times10^{33}$ & 0.2 & $9.7\times10^{4}$ & $3.9\times10^{11}$ & 4520 & $2\times10^{10}$ & 12 & $1.3\times10^{-14}$ \\\hline
J0605+3757 & Yes & $1.8\times10^{33}$ & 0.21 & $4.8\times10^{6}$ & $5.3\times10^{12}$ & 215 & $2\times10^{10}$ & 12 & $1.1\times10^{-14}$ \\\hline
J1545-4550 & No & $9.1\times10^{33}$ & 0.18 & $5.4\times10^{5}$ & $1.2\times10^{12}$ & 2222 & $2\times10^{10}$ & 11 & $9.8\times10^{-15}$ \\\hline
J0024-7204Q & No & $4.1\times10^{33}$ & 0.21 & $1.0\times10^{5}$ & $4.1\times10^{11}$ & 4520 & $2\times10^{10}$ & 12 & $9.6\times10^{-15}$ \\\hline
J1811-2405 & Yes & $5.8\times10^{33}$ & 0.27 & $5.4\times10^{5}$ & $1.3\times10^{12}$ & 1831 & $2\times10^{10}$ & 16 & $8.8\times10^{-15}$ \\\hline
J1653-2054 & No & $1.3\times10^{33}$ & 0.091 & $1.1\times10^{5}$ & $4.1\times10^{11}$ & 2631 & $2\times10^{10}$ & 5 & $8.7\times10^{-15}$ \\\hline
J0557+1550 & No & $3.5\times10^{33}$ & 0.23 & $4.2\times10^{5}$ & $1.0\times10^{12}$ & 1834 & $2\times10^{10}$ & 14 & $7.6\times10^{-15}$ \\\hline
J2022+2534 & No & $2.6\times10^{33}$ & 0.077 & $1.1\times10^{5}$ & $4.2\times10^{11}$ & 4015 & $2\times10^{10}$ & 5 & $7.4\times10^{-15}$ \\\hline
J1906+0055 & No & $1.2\times10^{33}$ & 0.13 & $5.3\times10^{4}$ & $2.6\times10^{11}$ & 4477 & $2\times10^{10}$ & 8 & $7.2\times10^{-15}$ \\\hline
J1142+0119 & Yes & $9.1\times10^{32}$ & 0.18 & $1.4\times10^{5}$ & $4.9\times10^{11}$ & 2169 & $2\times10^{10}$ & 11 & $6.4\times10^{-15}$ \\\hline
J0509+0856 & No & $5.2\times10^{32}$ & 0.13 & $4.2\times10^{5}$ & $1.0\times10^{12}$ & 817 & $2\times10^{10}$ & 8 & $5.8\times10^{-15}$ \\\hline
J1125-6014 & Yes & $1.6\times10^{33}$ & 0.33 & $7.6\times10^{5}$ & $1.6\times10^{12}$ & 988 & $2\times10^{10}$ & 20 & $5.3\times10^{-15}$ \\\hline
J1903-7051 & Yes & $1.8\times10^{33}$ & 0.34 & $9.5\times10^{5}$ & $1.8\times10^{12}$ & 930 & $2\times10^{10}$ & 20 & $4.8\times10^{-15}$ \\\hline
J2039-3616 & Yes & $1.9\times10^{33}$ & 0.16 & $5.0\times10^{5}$ & $1.2\times10^{12}$ & 1705 & $2\times10^{10}$ & 10 & $3.8\times10^{-15}$ \\\hline
J1405-4656 & No & $5.0\times10^{32}$ & 0.25 & $7.7\times10^{5}$ & $1.6\times10^{12}$ & 669 & $2\times10^{10}$ & 15 & $3.6\times10^{-15}$ \\\hline
J1630+3734 & Yes & $2.3\times10^{33}$ & 0.27 & $1.1\times10^{6}$ & $2.0\times10^{12}$ & 1187 & $2\times10^{10}$ & 16 & $3.3\times10^{-15}$ \\\hline
J0614-3329 & Yes & $4.4\times10^{33}$ & 0.32 & $4.6\times10^{6}$ & $5.3\times10^{12}$ & 630 & $2\times10^{10}$ & 19 & $3.2\times10^{-15}$ \\\hline
J2317+1439 & Yes & $4.7\times10^{32}$ & 0.2 & $2.1\times10^{5}$ & $6.6\times10^{11}$ & 1667 & $2\times10^{10}$ & 12 & $3.1\times10^{-15}$ \\\hline
J1641+3627F & No & $4.1\times10^{33}$ & 0.16 & $1.2\times10^{5}$ & $4.5\times10^{11}$ & 7420 & $2\times10^{10}$ & 9 & $3.0\times10^{-15}$ \\\hline
J0925+6103 & No & $6.0\times10^{32}$ & 0.15 & $2.1\times10^{5}$ & $6.5\times10^{11}$ & 1976 & $2\times10^{10}$ & 9 & $2.8\times10^{-15}$ \\\hline
J1741+1351 & Yes & $4.5\times10^{33}$ & 0.28 & $1.4\times10^{6}$ & $2.4\times10^{12}$ & 1667 & $2\times10^{10}$ & 17 & $2.3\times10^{-15}$ \\\hline
J1216-6410 & No & $2.9\times10^{32}$ & 0.18 & $3.5\times10^{5}$ & $9.2\times10^{11}$ & 1098 & $2\times10^{10}$ & 11 & $2.3\times10^{-15}$ \\\hline
J1017-7156 & No & $1.4\times10^{33}$ & 0.22 & $5.6\times10^{5}$ & $1.3\times10^{12}$ & 1807 & $2\times10^{10}$ & 13 & $2.1\times10^{-15}$ \\\hline
J2129-5721 & No & $3.2\times10^{33}$ & 0.15 & $5.7\times10^{5}$ & $1.3\times10^{12}$ & 3200 & $2\times10^{10}$ & 9 & $1.5\times10^{-15}$ \\\hline
J1804-2717 & No & $4.0\times10^{32}$ & 0.23 & $9.6\times10^{5}$ & $1.8\times10^{12}$ & 805 & $2\times10^{10}$ & 14 & $1.5\times10^{-15}$ \\\hline
J1732-5049 & Yes & $7.5\times10^{32}$ & 0.21 & $4.5\times10^{5}$ & $1.1\times10^{12}$ & 1874 & $2\times10^{10}$ & 12 & $1.4\times10^{-15}$ \\\hline
J1125+7819 & No & $7.4\times10^{32}$ & 0.33 & $1.3\times10^{6}$ & $2.3\times10^{12}$ & 903 & $2\times10^{10}$ & 20 & $1.4\times10^{-15}$ \\\hline
J1857+0943 & Yes & $9.1\times10^{32}$ & 0.28 & $1.1\times10^{6}$ & $2.0\times10^{12}$ & 1200 & $2\times10^{10}$ & 17 & $1.3\times10^{-15}$ \\\hline
J1921+1929 & Yes & $1.6\times10^{34}$ & 0.3 & $3.4\times10^{6}$ & $4.3\times10^{12}$ & 2434 & $2\times10^{10}$ & 18 & $1.2\times10^{-15}$ \\\hline
J2355+0051 & No & $4.9\times10^{32}$ & 0.28 & $1.0\times10^{6}$ & $1.9\times10^{12}$ & 958 & $2\times10^{10}$ & 17 & $1.2\times10^{-15}$ \\\hline
J1921+0137 & Yes & $9.8\times10^{33}$ & 0.27 & $8.6\times10^{5}$ & $1.7\times10^{12}$ & 5086 & $2\times10^{10}$ & 16 & $1.1\times10^{-15}$ \\\hline
J1125-5825 & Yes & $1.6\times10^{34}$ & 0.31 & $6.6\times10^{6}$ & $6.7\times10^{12}$ & 1744 & $2\times10^{10}$ & 18 & $9.5\times10^{-16}$ \\\hline
J1858-2216 & Yes & $2.2\times10^{33}$ & 0.25 & $4.0\times10^{6}$ & $4.7\times10^{12}$ & 921 & $2\times10^{10}$ & 15 & $9.4\times10^{-16}$ \\\hline
J1431-5740 & No & $7.3\times10^{32}$ & 0.18 & $2.4\times10^{5}$ & $7.1\times10^{11}$ & 3553 & $2\times10^{10}$ & 11 & $9.3\times10^{-16}$ \\\hline
J1918-0642 & No & $4.5\times10^{32}$ & 0.28 & $9.4\times10^{5}$ & $1.8\times10^{12}$ & 1111 & $2\times10^{10}$ & 17 & $9.0\times10^{-16}$ \\\hline
J1600-3053 & Yes & $1.6\times10^{33}$ & 0.24 & $1.2\times10^{6}$ & $2.2\times10^{12}$ & 1887 & $2\times10^{10}$ & 14 & $7.8\times10^{-16}$ \\\hline
J2015+0756 & No & $6.5\times10^{32}$ & 0.24 & $5.6\times10^{5}$ & $1.3\times10^{12}$ & 2086 & $2\times10^{10}$ & 14 & $7.4\times10^{-16}$ \\\hline
J0921-5202 & No & $1.5\times10^{32}$ & 0.27 & $3.3\times10^{6}$ & $4.2\times10^{12}$ & 355 & $2\times10^{10}$ & 16 & $5.4\times10^{-16}$ \\\hline
J2042+0246 & Yes & $1.2\times10^{33}$ & 0.22 & $6.7\times10^{6}$ & $6.6\times10^{12}$ & 640 & $2\times10^{10}$ & 13 & $5.3\times10^{-16}$ \\\hline
J1901+0300 & No & $7.6\times10^{32}$ & 0.16 & $2.1\times10^{5}$ & $6.5\times10^{11}$ & 5304 & $2\times10^{10}$ & 10 & $5.2\times10^{-16}$ \\\hline
J1137+7528 & Yes & $1.6\times10^{33}$ & 0.17 & $6.2\times10^{5}$ & $1.3\times10^{12}$ & 3833 & $2\times10^{10}$ & 10 & $4.9\times10^{-16}$ \\\hline
J1813-2621 & No & $1.1\times10^{33}$ & 0.22 & $7.1\times10^{5}$ & $1.5\times10^{12}$ & 3158 & $2\times10^{10}$ & 13 & $4.2\times10^{-16}$ \\\hline
J1813-0402 & No & $7.9\times10^{32}$ & 0.33 & $9.1\times10^{5}$ & $1.8\times10^{12}$ & 2229 & $2\times10^{10}$ & 20 & $4.0\times10^{-16}$ \\\hline
J1312+0051 & Yes & $1.8\times10^{33}$ & 0.21 & $3.3\times10^{6}$ & $4.1\times10^{12}$ & 1471 & $2\times10^{10}$ & 12 & $3.9\times10^{-16}$ \\\hline
J1841+0130 & No & $2.4\times10^{33}$ & 0.11 & $9.0\times10^{5}$ & $1.7\times10^{12}$ & 4230 & $2\times10^{10}$ & 7 & $3.7\times10^{-16}$ \\\hline
J0732+2314 & No & $6.9\times10^{32}$ & 0.17 & $2.6\times10^{6}$ & $3.5\times10^{12}$ & 1151 & $2\times10^{10}$ & 10 & $3.4\times10^{-16}$ \\\hline
J1120-3618 & No & $4.3\times10^{31}$ & 0.22 & $4.9\times10^{5}$ & $1.2\times10^{12}$ & 954 & $2\times10^{10}$ & 13 & $2.8\times10^{-16}$ \\\hline
J1929+0132 & No & $2.6\times10^{32}$ & 0.31 & $8.0\times10^{5}$ & $1.6\times10^{12}$ & 1852 & $2\times10^{10}$ & 19 & $2.2\times10^{-16}$ \\\hline
J1643-1224 & No & $1.5\times10^{33}$ & 0.14 & $1.3\times10^{7}$ & $1.0\times10^{13}$ & 740 & $2\times10^{10}$ & 8 & $2.2\times10^{-16}$ \\\hline
J1937+1658 & No & $1.7\times10^{33}$ & 0.22 & $1.5\times10^{6}$ & $2.4\times10^{12}$ & 3259 & $2\times10^{10}$ & 13 & $2.1\times10^{-16}$ \\\hline
J2001+0701 & No & $3.6\times10^{32}$ & 0.11 & $4.5\times10^{5}$ & $1.1\times10^{12}$ & 3782 & $2\times10^{10}$ & 7 & $1.7\times10^{-16}$ \\\hline
J1835-0114 & No & $4.1\times10^{32}$ & 0.21 & $5.8\times10^{5}$ & $1.3\times10^{12}$ & 3457 & $2\times10^{10}$ & 12 & $1.7\times10^{-16}$ \\\hline
J1455-3330 & Yes & $3.8\times10^{32}$ & 0.3 & $6.6\times10^{6}$ & $6.6\times10^{12}$ & 684 & $2\times10^{10}$ & 18 & $1.5\times10^{-16}$ \\\hline
J1751-2857 & No & $1.5\times10^{33}$ & 0.23 & $9.6\times10^{6}$ & $8.4\times10^{12}$ & 1087 & $2\times10^{10}$ & 13 & $1.4\times10^{-16}$ \\\hline
J1828+0625 & No & $7.8\times10^{32}$ & 0.32 & $6.7\times10^{6}$ & $6.8\times10^{12}$ & 1000 & $2\times10^{10}$ & 19 & $1.4\times10^{-16}$ \\\hline
J2019+2425 & No & $9.1\times10^{32}$ & 0.36 & $6.6\times10^{6}$ & $6.7\times10^{12}$ & 1163 & $2\times10^{10}$ & 22 & $1.2\times10^{-16}$ \\\hline
J2236-5527 & No & $2.3\times10^{32}$ & 0.26 & $1.1\times10^{6}$ & $2.0\times10^{12}$ & 2070 & $2\times10^{10}$ & 16 & $1.1\times10^{-16}$ \\\hline
J2302+4442 & Yes & $7.8\times10^{32}$ & 0.34 & $1.1\times10^{7}$ & $9.4\times10^{12}$ & 863 & $2\times10^{10}$ & 21 & $9.6\times10^{-17}$ \\\hline
J1908+0128 & No & $4.0\times10^{33}$ & 0.34 & $7.3\times10^{6}$ & $7.1\times10^{12}$ & 2627 & $2\times10^{10}$ & 20 & $9.1\times10^{-17}$ \\\hline
J1713+0747 & Yes & $7.1\times10^{32}$ & 0.32 & $5.9\times10^{6}$ & $6.2\times10^{12}$ & 1316 & $2\times10^{10}$ & 19 & $8.6\times10^{-17}$ \\\hline
J0203-0150 & No & $8.1\times10^{32}$ & 0.14 & $4.3\times10^{6}$ & $4.9\times10^{12}$ & 1837 & $2\times10^{10}$ & 9 & $8.1\times10^{-17}$ \\\hline
J1622-6617 & No & $3.5\times10^{31}$ & 0.11 & $1.4\times10^{5}$ & $5.0\times10^{11}$ & 4045 & $2\times10^{10}$ & 6 & $7.0\times10^{-17}$ \\\hline
J1709+2313 & No & $2.9\times10^{32}$ & 0.32 & $2.0\times10^{6}$ & $3.0\times10^{12}$ & 2179 & $2\times10^{10}$ & 19 & $5.5\times10^{-17}$ \\\hline
J1421-4409 & No & $3.7\times10^{32}$ & 0.21 & $2.7\times10^{6}$ & $3.6\times10^{12}$ & 2085 & $2\times10^{10}$ & 12 & $5.4\times10^{-17}$ \\\hline
J1811-0624 & No & $5.3\times10^{32}$ & 0.24 & $8.1\times10^{5}$ & $1.6\times10^{12}$ & 5801 & $2\times10^{10}$ & 14 & $4.8\times10^{-17}$ \\\hline
J2033+1734 & No & $4.2\times10^{32}$ & 0.22 & $4.9\times10^{6}$ & $5.4\times10^{12}$ & 1740 & $2\times10^{10}$ & 13 & $3.9\times10^{-17}$ \\\hline
J1900+0308 & No & $3.9\times10^{32}$ & 0.2 & $1.1\times10^{6}$ & $2.0\times10^{12}$ & 4801 & $2\times10^{10}$ & 12 & $3.6\times10^{-17}$ \\\hline
J1056-7117 & No & $2.7\times10^{31}$ & 0.15 & $7.9\times10^{5}$ & $1.6\times10^{12}$ & 1689 & $2\times10^{10}$ & 9 & $3.1\times10^{-17}$ \\\hline
J1853+1303 & No & $1.0\times10^{33}$ & 0.28 & $1.0\times10^{7}$ & $8.7\times10^{12}$ & 1887 & $2\times10^{10}$ & 17 & $3.0\times10^{-17}$ \\\hline
J1850+0124 & No & $1.9\times10^{33}$ & 0.29 & $7.3\times10^{6}$ & $7.1\times10^{12}$ & 3385 & $2\times10^{10}$ & 17 & $2.6\times10^{-17}$ \\\hline
J1640+2224 & Yes & $7.0\times10^{32}$ & 0.29 & $1.5\times10^{7}$ & $1.2\times10^{13}$ & 1370 & $2\times10^{10}$ & 17 & $2.2\times10^{-17}$ \\\hline
J1910+1256 & No & $6.2\times10^{32}$ & 0.22 & $5.1\times10^{6}$ & $5.5\times10^{12}$ & 2778 & $2\times10^{10}$ & 13 & $2.1\times10^{-17}$ \\\hline
J1824-0621 & Yes & $2.1\times10^{33}$ & 0.34 & $8.7\times10^{6}$ & $8.1\times10^{12}$ & 3534 & $2\times10^{10}$ & 20 & $2.1\times10^{-17}$ \\\hline
J2229+2643 & No & $4.6\times10^{32}$ & 0.14 & $8.0\times10^{6}$ & $7.4\times10^{12}$ & 1800 & $2\times10^{10}$ & 8 & $2.1\times10^{-17}$ \\\hline
J1844+0115 & No & $1.2\times10^{33}$ & 0.16 & $4.4\times10^{6}$ & $4.9\times10^{12}$ & 4358 & $2\times10^{10}$ & 10 & $2.0\times10^{-17}$ \\\hline
J1806+2819 & No & $8.6\times10^{31}$ & 0.29 & $3.8\times10^{6}$ & $4.6\times10^{12}$ & 1330 & $2\times10^{10}$ & 17 & $1.9\times10^{-17}$ \\\hline
J1855-1436 & Yes & $1.9\times10^{33}$ & 0.31 & $5.3\times10^{6}$ & $5.8\times10^{12}$ & 5127 & $2\times10^{10}$ & 19 & $1.7\times10^{-17}$ \\\hline
J1824+1014 & Yes & $6.3\times10^{32}$ & 0.31 & $7.1\times10^{6}$ & $7.0\times10^{12}$ & 2904 & $2\times10^{10}$ & 18 & $1.2\times10^{-17}$ \\\hline
J1938+2012 & No & $3.2\times10^{32}$ & 0.21 & $1.4\times10^{6}$ & $2.3\times10^{12}$ & 6301 & $2\times10^{10}$ & 12 & $1.2\times10^{-17}$ \\\hline
J1825-0319 & No & $5.7\times10^{32}$ & 0.21 & $4.5\times10^{6}$ & $5.1\times10^{12}$ & 3862 & $2\times10^{10}$ & 12 & $1.2\times10^{-17}$ \\\hline
J2010+3051 & No & $3.4\times10^{32}$ & 0.24 & $2.0\times10^{6}$ & $3.0\times10^{12}$ & 6450 & $2\times10^{10}$ & 14 & $7.4\times10^{-18}$ \\\hline
J1708-3506 & No & $9.9\times10^{32}$ & 0.19 & $1.3\times10^{7}$ & $1.0\times10^{13}$ & 3320 & $2\times10^{10}$ & 11 & $6.9\times10^{-18}$ \\\hline
J1930+2441 & No & $3.6\times10^{32}$ & 0.27 & $6.6\times10^{6}$ & $6.6\times10^{12}$ & 3267 & $2\times10^{10}$ & 16 & $6.1\times10^{-18}$ \\\hline
J1810-2005 & No & $3.3\times10^{31}$ & 0.33 & $1.3\times10^{6}$ & $2.3\times10^{12}$ & 3513 & $2\times10^{10}$ & 20 & $4.2\times10^{-18}$ \\\hline
J1327-0755 & Yes & $7.4\times10^{32}$ & 0.26 & $7.3\times10^{5}$ & $1.5\times10^{12}$ & 25000 & $2\times10^{10}$ & 16 & $4.1\times10^{-18}$ \\\hline
J1913+0618 & No & $6.0\times10^{32}$ & 0.33 & $5.9\times10^{6}$ & $6.2\times10^{12}$ & 5862 & $2\times10^{10}$ & 20 & $3.7\times10^{-18}$ \\\hline
J1955+2908 & No & $1.0\times10^{33}$ & 0.21 & $1.0\times10^{7}$ & $8.7\times10^{12}$ & 6303 & $2\times10^{10}$ & 12 & $2.7\times10^{-18}$ \\\hline
J1529-3828 & No & $3.5\times10^{32}$ & 0.19 & $1.0\times10^{7}$ & $8.8\times10^{12}$ & 4301 & $2\times10^{10}$ & 11 & $1.9\times10^{-18}$ \\\hline
J0407+1607 & No & $3.7\times10^{31}$ & 0.22 & $5.8\times10^{7}$ & $2.8\times10^{13}$ & 1337 & $2\times10^{10}$ & 13 & $2.1\times10^{-19}$ \\\hline
J1912-0952 & No & $1.4\times10^{31}$ & 0.085 & $2.5\times10^{6}$ & $3.4\times10^{12}$ & 9446 & $2\times10^{10}$ & 5 & $1.1\times10^{-19}$ \\\hline
\caption{\label{tab:real Hes} The line flux estimates for He WD-MSP systems in the ATNF Pulsar Catalogue. The values for companion mass and distance are taken from the ATNF Pulsar Catalogue \citep{Manchester_2005}. The companion mass used here is the median mass according to the ATNF Pulsar Catalogue. The separation is calculated from the orbit period listed in the ATNF Pulsar Catalogue. The radius is the lesser value of either the size of the Roche lobe or the radius of the model used in the Methods section. The density column is the average density assuming a homogeneous sphere with the companion mass and radius.}
\end{longtable*}
\clearpage


\bibliography{refs}{}
\bibliographystyle{aasjournalv7}




\end{document}